\documentclass[12pt]{article}
\newcommand{\beq}{\begin{equation}}
\newcommand{\eeq}{\end{equation}}
\newcommand{\beqa}{\begin{eqnarray}}
\newcommand{\eeqa}{\end{eqnarray}}
\newcommand{\beqar}{\begin{eqnarray*}}
\newcommand{\eeqar}{\end{eqnarray*}}

\newcommand{\eps}{\epsilon}
\newcommand{\ga}{\gamma}
\newcommand{\Ga}{\Gamma}

\newcommand{\inn}{\!\cdot\!}

\newcommand{\z}{\zeta}

\newcommand{\eg}{{\it e.g.,}\ }
\newcommand{\ie}{{\it i.e.,}\ }
\newcommand{\labell}[1]{\label{#1}} 
\newcommand{\reef}[1]{(\ref{#1})}
\newcommand\prt{\partial}
\newcommand\veps{\varepsilon}

\newcommand\cF{{\cal F}}

\newcommand\cA{{\cal A}}

\newcommand\cM{{\cal M}}

\newcommand\cB{{\cal B}}

\newcommand\bz{\bar{z}}

\newcommand\ti{{\tilde  i}}
\newcommand\tj{{\tilde  j}}

\newcommand\Tr{{\rm Tr}}

\begin{document}

\vspace*{1cm}

\begin{center}
{\bf \Large   S-matrix element of two R-R and one NS states }

\vspace*{1cm}

{Mojtaba Mohammadzadeh\footnote{Mohammadzadeh.mojtaba@mail.um.ac.ir}   and 
 Mohammad R. Garousi\footnote{garousi@um.ac.ir} }\\
\vspace*{1cm}
{ Department of Physics, Faculty of Science, Ferdowsi University of Mashhad,\\ P.O. Box 1436, Mashhad, Iran}
\\
\vspace{2cm}

\end{center}

\baselineskip=18pt

\begin{abstract}

We explicitly calculate the disk-level S-matrix element of two closed string R-R and one open string NS vertex operators in RNS formalism. We show that the amplitude  satisfies various duality Ward identities. In particular, when one of the R-R is zero-form, the other one is two-form and the NS state is gauge boson, the amplitude transforms under S-duality Ward identity  to the   amplitude of one dilaton, one B-field and one gauge boson which has been recently calculated explicitly.

We have also proposed a  soft theorem for disk-level scattering amplitude of arbitrary number of hard closed strings and one soft open string    at the leading order of soft momentum, and shown that the above amplitude satisfies the soft theorem.


\end{abstract}
\newpage
\section{Introduction}

Perturbative spectrum of  type II closed superstring   in flat spacetime consists of a tower of bosonic states in NS-NS and R-R sectors, and their corresponding fermionic states in   the R-NS and NS-R sectors (see \eg \cite{Becker:2007zj}). Non-perturbative spectrum of the type II superstring theory includes dynamical  D$_p$-branes objects \cite{Polchinski:1995mt}. The perturbative excitations of the D$_p$-branes are given by open string spectrum which consists of bosonic states in NS sector and their corresponding fermionic states in the R sector. In perturbative theory, the leading interaction of excited D$_p$-brane with the closed string states are given by the S-matrix elements of the corresponding closed and open string vertex operators on the disk world-sheet \cite{Garousi:1996ad,Hashimoto:1996kf}. The type II theory has various   dualities \cite{Giveon:1994fu,Schwarz:1996bh} which appear in the S-matrix elements   through the  corresponding Ward identities \cite{Garousi:2011we,Garousi:2017fbe}.

 A duality of type II theory is T-duality which appears when one considers the theory on a compact manifold. In the simplest case that the compact manifold is a circle,   the closed string  spectrum of type IIA theory on the circle with radius R transforms under T-duality to the closed string  spectrum of type IIB on a circle with radius $\alpha'/R$. Moreover, the D$_p$-brane along the circle in type IIA theory transforms to D$_{p-1}$-brane orthogonal to the dual circle in the type IIB theory.  The T-duality Ward identity indicates that the  disk-level S-matrix elements on the world-volume of D$_p$-brane in  type IIA theory on a circle transforms under linear T-duality to the corresponding disk-level S-matrix elements on the world-volume of D$_{p-1}$-brane. The T-duality Ward identity has been used in \cite{Velni:2012sv,Velni:2013jha,Jalali:2016xtv} to generate various disk-level S-matrix elements.

The type IIB theory enjoys also   S-duality  which indicates that the spectrum of type IIB in flat spacetime transforms covariantly  under $SL(2,R)$ transformation. In particular, the D$_3$-brane is invariant under the $SL(2,R)$ transformation, the NS-NS antisymmetric B-field and R-R two-form transforms as doublet under under the $SL(2,R)$ transformation. The S-duality Ward identity indicates that the  disk-level S-matrix elements on the world-volume of D$_3$-brane in  type IIB   transforms under linear $SL(2,R)$ transformation to the corresponding disk-level S-matrix elements on the world-volume of D$_3$-brane. This Ward identity may be used to generate the complicated  S-matrix elements of R-R vertex operators which involve spin operator \cite{Cohn:1986bn} from the corresponding S-matrix elements of  NS-NS vertex operators which are straightforward to calculate them. 

As an example of the S-matrix elements of the R-R vertex operators, in this paper, we explicitly calculate the disk-level S-matrix element of two R-R and one NS vertex operators in RNS formalism. Such amplitude has been recently predicted by the S-duality Ward identity \cite{Garousi:2012gh}. We observe that the explicit  calculations produce exactly the   amplitude predicted by the S-duality. The S-duality Ward identity indicates that apart from the overall dilaton factor of background dilaton, the   disk-level S-matrix elements must combine into S-dual multiplets which are invariant under the linear $SL(2,R)$ transformation \cite{Garousi:2012gh}. This indicates that the amplitudes involving two R-R and one NS states which can not be written in S-dual multiplet,  must be zero. We observe  that the amplitudes that are predicted by S-duality to be zero, \eg $C^{(0)}C^{(0)}F$-amplitude, are in fact zero by explicit calculation.

A consistency check of the S-matrix elements in string theory is that they must satisfy the soft graviton/photon theorems \cite{Ademollo:1975pf}-\cite{DiVecchia:2017gfi} in which one graviton/photon is soft. In the soft theorems \cite{Ademollo:1975pf}-\cite{DiVecchia:2017gfi}, however, the external states are all either closed string states or all open string states in which we are not interested in this paper. When one string state is soft open string and all other states are hard closed string stats, one can easily find the corresponding soft therm at the leading order of the soft open string momentum. 

The Ward identity corresponding to the gauge boson transformation indicates that the disk-level S-matrix element  of $n$ closed strings and one open string gauge field must be in  the following form\footnote{Using conservation of momentum along the D-brane, \ie $2k+ p_1+p_1\inn D+ p_2+ p_2\inn D+ \cdots+ p_n+p_n\inn D=0 $, one may write $p_n\inn D$ in terms of other momenta.}:
\beqa
\cA_{n+1}&=&f^{ab}[\cA_n(k, p_1,p_1\inn D, p_2, p_2\inn D, \cdots, p_n )]_{ab} \nonumber
\eeqa
where $f^{ab}$ is the gauge field strength in momentum space and  $(\cA_n)_{[ab]}$ is the factor which does not involve the open string gauge field polarization\footnote{Our index convention is that the Greek letters  $(\mu,\nu,\cdots)$ are  the indices of the space-time coordinates, the Latin letters $(a,d,c,\cdots)$ are the world-volume indices and the letters $(i,j,k,\cdots)$ are the normal bundle indices.}. In general, this factor is a complicated function of the momentum of the gauge field $k^a$. However, using the fact that there is only one open string state, one observes that there is no pole $1/k\cdot k$ in the amplitude. As a result, when the gauge field is soft, \ie $k^a\rightarrow 0$, one finds the above amplitude defines a soft theorem which involves   the soft factor $f^{ab}$ at the leading order, and the hard factor  $[\cA_n]_{ab}$ which involves only polarizations and the momenta of the $n$ closed string states, \ie

\beqa
\cA_{n+1}&=&f^{ab}[\cA_n(p_1,p_1\inn D, p_2, p_2\inn D, \cdots, p_n )]_{ab}\labell{A1} 
\eeqa
 The trace of  $[\cA_n]_{ab}$ is the disk-level scattering amplitude of $n$ closed strings which  is zero because $[\cA_n]_{ab}$ is antisymmetric. The above relation exists at any order of $\alpha'$, so one expects the disk-level S-matrix elements satisfy this theorem  for the soft gauge field. 

There is similar theorem when the open string state is transverse scalar field. To find such theorem at the leading order of the scalar momentum,  we consider the observation that the closed string fields in effective world-volume action must be the Taylor expansion of the transverse scalar fields \cite{Garousi:1998fg}, \ie $C(\Phi^i)=C+\Phi^i\prt_i C+\cdots$ where $C$ is a closed string field. Using  the  coupling $\Phi^i\prt_i C$, one can easily write the S-matrix element of $n$ closed string and one transverse scalar field at  the leading order of $k^a$ and at leading order of $\alpha'$ to be    
\beqa
\cA_{n+1}&=&\z^i (p_1+p_2+\cdots +p_n)_i\cA_n \labell{A2}
\eeqa
where $\z^i$ is polarization of the scalar fields and $\cA_n$ is the scattering amplitude of $n$ closed string states at low energy. The above relation must be valid for any order of $\alpha'$. As a result, the disk-level S-matrix element of one transverse scalar and $n$ closed string states must satisfy the above soft theorem. We will show that the scattering amplitude of two R-R and one NS states satisfies the above soft theorems.

 An outline of the paper is as follows: We begin the section 2 by explicitly calculating the disk-level scattering amplitude of two R-R and one NS vertex operators in RNS formalism. We use  $(-1/2,-1/2)$-picture for the R-R vertex operators. In this picture, the field strengths of the R-R fields appear in the vertex operators, as a result, the amplitude satisfies the R-R gauge symmetry Ward identity from the onset. We show that the final amplitude satisfies the open string gauge symmetry Ward identity as well. In section 3, we show that the amplitude satisfies the S-duality Ward identity. In particular, the amplitude of one R-R zero-form, one R-R two form and one NS gauge field transforms under the S-duality Ward identity to the amplitude of one NS-NS dilaton, one B-field and one NS gauge field that has been recently calculated explicitly. In this section, we have also shown that the amplitudes that are predicted by the S-duality Ward identity to be zero, are in fact zero. In section 4, we explicitly write the D$_p$-brane  amplitudes that are non-zero for   $p=0,1,2,3$. In section 5, we show that these amplitude satisfies the T-duality Ward identity. In section 6, we show that the amplitudes that we have found in section 4, satisfy the soft scalar theorem \reef{A2}.  In section 7, we show that the amplitudes satisfy the soft-photon theorem and find the kinematic factors in $[\cA_2]_{ab}$.

 \section{Explicit calculation of the amplitude}

The tree-level scattering amplitude of two R-R closed string and one NS open string states  on the world-volume of a D$_p$
-brane is given by the correlation function of their corresponding vertex operators on disk (see \eg \cite{Garousi:2017fbe}).
Since the background charge of the world-sheet with topology of a disk is $Q_{\phi}= 2$, one has
to choose the vertex operators in the appropriate pictures to produce the compensating
charge
$Q_{\phi}= -2$. The scattering amplitude may then be given by the following correlation
function:
 \beqa
\cA(\veps_1,p_1;\veps_2,p_2;\z_3,k_3)&\sim&<V_{RR}^{(-1/2,-1/2)}(\veps_1,p_1)V_{RR}^{(-1/2,-1/2)}(\veps_2,p_2)V_{NS}^{(0)}(\z_3,k_3)>\labell{amp23}
\eeqa
where the vertex operators, after using the doubling trick \cite{Garousi:1996ad}, are
\beqa
V_{RR}^{(-1/2,-1/2)}&\!\!\!\!\!=\!\!\!\!\!&(P_-\Gamma_{1(n)}M_p)^{AB}\int d^2z_1:e^{-\phi(z_1)/2}S_A(z_1)e^{ip_1\cdot X}:e^{-\phi(\bz_1)/2}S_B(\bz_1)e^{ip_1\cdot D\cdot  X}:\\\nonumber
V_{RR}^{(-1/2,-1/2)}&\!\!\!\!\!=\!\!\!\!\!&(P_-\Gamma_{2(m)}M_p)^{CD}\int d^2z_2:e^{-\phi(z_2)/2}S_C(z_2)e^{ip_2\cdot X}:e^{-\phi(\bz_2)/2}S_D(\bz_2)e^{ip_2\cdot D\cdot  X}:\\\nonumber
V_{NS}^{(0)}&\!\!\!\!\!=\!\!\!\!\!&(\z_3)_\mu \int dz_3:(\prt X^{\mu}+2ik_3\inn\psi\psi^{\mu})e^{2ik_3\cdot X}
\eeqa
where $z_3$ is along the real axis and $z_1,z_2$ are in upper-half  $z$-plane. The index $\mu$ is the world volume indice $a$ when the NS state is a gauge boson and is the transverse index $i$ when the NS state is the transverse scalar fields. The indices $A,B,\cdots$ are the Dirac spinor indices and  $P_-=\frac{1}{2}(1-\gamma_{11})$ is the chiral projection operator which makes the calculation of the gamma matrices to be with the full $32\times 32$ Dirac matrices of the ten dimensions. The  R-R polarization appears in  $\Gamma_{(n)}$ and the world-volume Levi-Civita tensor  appears in $M_p$, \ie
\beqa
\Gamma_{(n)}&=&\frac{1}{n!}F_{\mu_1\cdots\mu_{n}}\gamma^{\mu_1}\cdots\gamma^{\mu_{n}}\nonumber\\
M_p&=&\frac{\pm 1}{(p+1)!} \eps_{a_0 \cdots a_p} \ga^{a_0} \cdots \ga^{a_p}
\eeqa
where $F^{(n)}$ is the linearized field strength of the R-R potential $C^{(n-1)}$. The matrix $D_{\mu\nu}$ is a diagonal matrix which is the world-volume flat metric when the indices are the world volume indices, and is minus the flat metric of the transverse space when the indices are the transverse indices. We have used the integral form for all vertex operators, as a result, the integrand of the amplitude must be invariant under the conformal transformation of the upper-half plane which is $SL(2,R)$ transformation.

Using the standard upper-half $z$-plane  propagators
\beqa
<X^{\mu}(x)X^{\nu}(y)>&=&-\eta^{\mu\nu}\log(x-y)\nonumber\\
<\phi(x)\phi(y)>&=&-\log(x-y)\labell{wpro}
\eeqa
one can easily calculate the ghost $\phi$ and $X^{\mu}$ correlators in \reef{amp23}. To calculate the correlation functions between the spin operators and the world-sheet field $\psi^{\mu}$, we use the following operator product expansion:
\beqa
:S_A(z_1):\psi^\mu\psi^\nu(z_2): =-\frac{(\Gamma^{\mu\nu})_{A}^{}{}^{E}}{2z_{21}}:S_E(z_1):
\eeqa
where $z_{ij}=z_i-z_j$ and $\Gamma^{\mu\nu}$ is $\frac{1}{2}(\gamma^{\mu}\gamma^{\nu}-\gamma^{\nu}\gamma^{\mu})$, to reduce the correlators to the following  known correlation function \cite{Cohn:1986bn}:
\beqa
<:S_A(z_1):S_B(z_2):S_C(z_3):S_D(z_4):> = I_{ABCD}
\eeqa
where
\beqa
I_{ABCD} = \frac {z_{14} z_{23}(\gamma^\mu)_{AB} (\gamma_\mu)_{CD} - z_{12} z_{34} (\gamma^\mu)_{AD} (\gamma_\mu)_{BC}}{2(z_{12} z_{13} z_{14} z_{23} z_{24} z_{34})^\frac{3}{4}}
\eeqa
This calculation produces the  following 14 terms:
\beqa
\cA(\veps_1,p_1;\veps_2,p_2;\z_3,k_3)&\sim& \int d^2z_1d^2z_2dz_3 \,K\sum_{i=1}^{14}A_i\labell{amp0}
\eeqa
where $A_1,\cdots, A_{14}$ in the integrand are the following:
\beqa
A_1&=&\frac{i (P_-\Gamma_{1(n)}M_p)^{AB}(\gamma^\mu)_{AB}(P_-\Gamma_{2(m)}M_p)^{CD}(\gamma_\mu)_{CD}(\z_3)^{\chi}p_2 \cdot D_{\chi} }{z_{12}z_{1\bar{1}}z_{2\bar{2}}z_{3\bar{1}}z_{3\bar{2}}}\\
A_2&=&-\frac{i (P_-\Gamma_{1(n)}M_p)^{AB}(\gamma_\mu)_{BC}(P_-\Gamma_{2(m)}M_p)^{CD}(\gamma^\mu)_{AD}(\z_3)^{\chi}p_2 \cdot D_{\chi} }{z_{12}z_{1\bar{2}}z_{3\bar{1}}z_{3\bar{2}}z_{\bar{1}2}}\nonumber\\
A_3&=&-\frac{i(P_-\Gamma_{1(n)}M_p)^{AB}(\Gamma^{b\chi})_A^{}{}^E (\gamma^\mu)_{EB}  \ (k_3)_b\ (\z_3)_{\chi}(P_-\Gamma_{2(m)}M_p)^{CD}(\gamma_\mu)_{CD}}{z_{12}z_{1\bar{1}}z_{2\bar{2}}z_{31}z_{\bar{1}\bar{2}}}\nonumber\\
A_4&=&-\frac{i (P_-\Gamma_{1(n)}M_p)^{AB}(\gamma^\mu)_{AB}  \ (k_3)_b\ (\z_3)_{\chi}(P_-\Gamma_{2(m)}M_p)^{CD}(\Gamma^{b\chi})_C^{}{}^E(\gamma^\mu)_{ED}}{z_{12}z_{1\bar{1}}z_{2\bar{2}}z_{32}z_{\bar{1}\bar{2}}}\nonumber\\
A_5&=&-\frac{i(P_-\Gamma_{1(n)}M_p)^{AB}(\Gamma^{b\chi})_B^{}{}^E (\gamma^\mu)_{AE}  \ (k_3)_b\ (\z_3)_{\chi}(P_-\Gamma_{2(m)}M_p)^{CD}(\gamma_\mu)_{CD}}{z_{12}z_{1\bar{1}}z_{2\bar{2}}z_{3\bar{1}}z_{\bar{3}\bar{2}}}\nonumber\\
A_6&=&-\frac{i (P_-\Gamma_{1(n)}M_p)^{AB}(\gamma^\mu)_{AB}(P_-\Gamma_{2(m)}M_p)^{CD}(\gamma_\mu)_{CD}(\z_3)^{\chi}p_{1\chi} }{z_{12}z_{2\bar{2}}z_{31}z_{3\bar{1}}z_{\bar{1}\bar{2}}}\nonumber\\
A_7&=&-\frac{i (P_-\Gamma_{1(n)}M_p)^{AB}(\gamma_\mu)_{BC}(P_-\Gamma_{2(m)}M_p)^{CD}(\gamma^\mu)_{AD}(\z_3)^{\chi}p_{2\chi} }{z_{12}z_{1\bar{2}}z_{32}z_{3\bar{1}}z_{\bar{1}\bar{2}}}\nonumber\\
A_8&=&-\frac{i (P_-\Gamma_{1(n)}M_p)^{AB}(\gamma^\mu)_{AB}(P_-\Gamma_{2(m)}M_p)^{CD} (k_3)_b(\gamma_\mu)_{CE}(\Gamma^{b\chi})_D^{}{}^E(\z_3)_{\chi}}{z_{12}z_{1\bar{1}}z_{2\bar{2}}z_{3\bar{2}}z_{\bar{1}\bar{2}}}\nonumber\\
A_9&=&\frac{i(P_-\Gamma_{1(n)}M_p)^{AB}(\Gamma^{a\chi})_A^{}{}^E (\gamma^\mu)_{ED}  \ (k_3)_a\ (\z_3)_{\chi}(P_-\Gamma_{2(m)}M_p)^{CD}(\gamma_\mu)_{BC}}{z_{12}z_{1\bar{2}}z_{31}z_{\bar{1}2}z_{\bar{1}\bar{2}}}\nonumber\\
A_{10}&=&\frac{i (P_- \Gamma_{1(n)} M_p)^{AB} (\Gamma^{a\chi})_C^{}{}^E (\gamma^\mu)_{AD}  \ 
(k_3)_a\ (\z_3)_{\chi}(P_- \Gamma_{2(m)}M_p)^{CD}(\gamma_\mu)_{BE}}{z_{12}
 z_{1\bar{2}} z_{32} z_{\bar{1}2} z_{\bar{1}\bar{2}}} \nonumber\\
A_{11}&=&\frac{i (P_-\Gamma_{1(n)}M_p)^{AB}(\Gamma^{a\chi})_B^{}{}^E (\gamma^\mu)_{AD}  \ (k_3)_a\ (\z_3)_{\chi}(P_-\Gamma_{2(m)}M_p)^{CD}(\gamma_\mu)_{EC}}{z_{12}z_{1\bar{2}}z_{3\bar{1}}z_{\bar{1}2}z_{\bar{1}\bar{2}}}\nonumber\\
A_{12}&=&\frac{i z_{1\bar{1}} (P_-\Gamma_{1(n)}M_p)^{AB}(\gamma_\mu)_{BC}(P_-\Gamma_{2(m)}M_p)^{CD}(\gamma^\mu)_{AD}(\z_3)^{\chi}p_{1\chi}}{ {z_{12}z_{1\bar{2}}z_{31}z_{3\bar{1}}z_{\bar{1}2}z_{\bar{1}\bar{2}}}}\nonumber\\
A_{13}&=&\frac{i (P_-\Gamma_{1(n)}M_p)^{AB}(\Gamma^{a\chi})_D^{}{}^E (\gamma^\mu)_{AE}  \ (k_3)_a\ (\z_3)_{\chi}(P_-\Gamma_{2(m)}M_p)^{CD}(\gamma_\mu)_{BC}}{z_{12}z_{1\bar{2}}z_{3\bar{2}}z_{\bar{1}2}z_{\bar{1}\bar{2}}}\nonumber\\
A_{14}&=&\frac{i z_{\bar{1}2} (P_-\Gamma_{1(n)}M_p)^{AB}(\gamma^\mu)_{AB}(P_-\Gamma_{2(m)}M_p)^{CD}(\gamma_\mu)_{CD}(\z_3)^{\chi}p_{2\chi} }{z_{12}z_{1\bar{1}}z_{2\bar{2}}z_{32}z_{3\bar{1}}z_{\bar{1}\bar{2}}}\nonumber
\eeqa
and the overall kinematic factor $K$ is 
\beqa
K&=&z_{ 1\bar{1}}^{p_1.D.p_1}|z_{12}|^{2p_1.p_2}|z_{ 1\bar{2}}|^{2p_1.D.p_2}|z_{13}|^{4p_1.k_3}z_{ 2\bar{2}}^{p_2.D.p_2}|z_{23}|^{4p_2.k_3}
\eeqa
One can easily verify that each integrand is invariant under $SL(2,R)$ transformation. 

For subsequent discussions, we rewrite the amplitude as 
\beqa
\cA(\veps_1,p_1;\veps_2,p_2;\z_3,k_3)&\sim&\sum_{i=1}^{14}a_i  q_i\labell{amp}
\eeqa
where  $q_i$'s are integrals of some function of $z_i$'s and momenta, \eg
\beqa
q_1&=&\int d^2z_1d^2z_2dz_3 \frac{K}{z_{12}z_{1\bar{1}}z_{2\bar{2}}z_{3\bar{1}}z_{3\bar{2}}}
\eeqa
and  $a_i$'s are the remaining terms in $A_i$'s, \eg
\beqa
a_1&=&i (P_-\Gamma_{1(n)}M_p)^{AB}(\gamma^\mu)_{AB}(P_-\Gamma_{2(m)}M_p)^{CD}(\gamma_\mu)_{CD}(\z_3)^{\chi}p_2 \cdot D_{\chi} 
\eeqa
Contracting the Dirac indices, one finds each of $a_2,\,a_7,\,a_9,\,a_{10},\,a_{11},\,a_{12},\,a_{13}$ produces   one trace over the gamma matrices, \ie
\beqa
&&a_2=ib_2(\z_3)^{\chi}p_2 \cdot D_{\chi},\,\,  
a_7=ib_2(\z_3)^{\chi}p_{2\chi},\,\,
a_{12}=ib_2(\z_3)^{\chi}p_{1\chi}\nonumber\\
&&a_9= ib_9^{a\chi}(k_3)_a(\z_3)_{\chi},\,\,
 a_{10}=a_9(1\leftrightarrow 2) \nonumber\\
&&a_{13}=-i b_{13}^{a\chi}(k_3)_a(\z_3)_{\chi},\,\,
a_{11}=a_{13}(1\leftrightarrow 2) \labell{rel5}
\eeqa
where $b_2,b_9,b_{13}$ are the following one-trace terms:
\beqa
b_2&=&\Tr(P_-\Gamma_{1(n)}M_p\gamma_\mu \Gamma_{2(m)}M_p\gamma^\mu)\\
b_9^{a\chi}&=&\Tr(P_-\Gamma_{1(n)}M_p\gamma_\mu \Gamma_{2(m)}M_p\gamma^\mu\Gamma^{a\chi}) \\
b_{13}^{a\chi}&=&\Tr(P_-\Gamma_{1(n)}M_p\gamma_\mu \Gamma_{2(m)}M_p\Gamma^{a\chi}\gamma^\mu) \labell{rel4}
\eeqa
One also finds that each of $a_1,\,a_3,\,a_4,\,a_5,\,a_6,\,a_8,\,a_{14}$ produces  two traces over the gamma matrices,\ie 
\beqa
&&a_1=ib_1(\z_3)^{\chi}p_2 \cdot D_{\chi},\,\,  
a_6=ib_1(\z_3)^{\chi}p_{1\chi},\,\,
a_{14}=ib_1(\z_3)^{\chi}p_{2\chi}\nonumber\\
&&a_3= ib_3^{b\chi} \ (k_3)_b\ (\z_3)_{\chi},\,\,
a_{4}=a_3(1\leftrightarrow 2) \nonumber\\
&&a_{5}=-i b_{5}^{b\chi}(k_3)_a\ (\z_3)_{\chi},\,\,
a_{8}=b_{5}(1\leftrightarrow 2) \nonumber
\eeqa
where $b_1,b_3,b_{5}$ are the following two-trace terms:
\beqa
b_1&=&\Tr (P_-\Gamma_{1(n)}M_p\gamma^\mu) \Tr(P_-\Gamma_{2(m)}M_p\gamma_\mu)\nonumber\\
b_3^{b\chi}&=&\Tr(P_-\Gamma_{1(n)}M_p\gamma^\mu\Gamma^{b\chi}) \Tr (P_-\Gamma_{2(m)}M_p\gamma_\mu)\nonumber \\
b_{5}^{b\chi}&=&\Tr(P_-\Gamma_{1(n)}M_p\Gamma^{b\chi}\gamma^\mu)  \Tr(P_-\Gamma_{2(m)}M_p\gamma_\mu) \labell{rel1}
\eeqa
Each trace in $b_1,b_3^{b\chi},b_5^{b\chi}$ has one factor of $M_p$, whereas  in $b_2,b_9^{a\chi},b_{13}^{a\chi}$ each trace has two factors of $M_p$. This makes the calculation of the traces in $b_2,b_9^{a\chi},b_{13}^{a\chi}$ to be difficult for general D$_p$-brane. Using the Gamma package \cite{Gran:2001yh}, we have found the following relations:
\beqa
b_2&=&c_2-b_1 \nonumber\\
b_9^{a\chi}&=&c_9^{a\chi}-b_3^{a\chi}\nonumber\\
b_{13}^{a\chi}&=&c_{13}^{a\chi}-b_5^{a\chi}\labell{rel}
\eeqa
where $c_2,c_9^{a\chi},c_{11}^{a\chi}$ are the following one-trace terms:
\beqa
c_{2}&\equiv&\Tr(P_-\Gamma_{1(n)}D_\mu{}{}^\nu \gamma_\nu M_p C^{-1}M_p^T\Gamma_{2(m)}^T C\gamma^\mu) \nonumber\\
c_9^{a\chi}&\equiv&\Tr(P_-\Gamma_{1(n)}D_\mu{}{}^\nu \gamma_\nu M_p C^{-1}M_p^T\Gamma_{2(m)}^T C\gamma^\mu\Gamma^{a\chi})\nonumber\\
c_{13}^{a\chi}&\equiv&\Tr(P_-\Gamma_{1(n)}D_\mu{}{}^\nu \gamma_\nu M_p C^{-1}M_p^T\Gamma_{2(m)}^T C\Gamma^{a\chi}\gamma^\mu)\labell{rel20}
\eeqa
The first relation in \reef{rel} has been found in \cite{Garousi:1996yt} by using the KLT relation that maps  the amplitude of  four open string fermions to the amplitude of two closed string RR bosons. This relation has been found by using two different amplitudes for four fermions that are produced by different ways of fixing the $SL(2,R)$ symmetry of the disk amplitude.  Now using the relation $M_p C^{-1}M_p^TC =1$ \cite{Garousi:1996ad}, one finds the traces in $c_2,c_9^{a\chi},c_{13}^{a\chi}$ have in fact no factor of $M_p$, \ie
\beqa
c_2&=&\Tr(P_-\Gamma_{1(n)}D_\mu{}{}^\nu \gamma_\nu \Gamma_{2(m)} \gamma^\mu)(-1)^{\frac{1}{2}m(m+1)}\nonumber\\
c_9^{a\chi}&=&\Tr(P_-\Gamma_{1(n)}D_\mu{}{}^\nu \gamma_\nu \Gamma_{2(m)} \gamma^\mu \Gamma^{a\chi})(-1)^{\frac{1}{2}m(m+1)}\nonumber\\
c_{13}^{a\chi}&=&\Tr(P_-\Gamma_{1(n)}D_\mu{}{}^\nu \gamma_\nu \Gamma_{2(m)} \Gamma^{a\chi}\gamma^\mu )(-1)^{\frac{1}{2}m(m+1)}\labell{rel2}
\eeqa
So $c_2,c_9^{a\chi},c_{13}^{a\chi}$ are independent of the dimension of D$_{p}$-brane. One has to perform the traces in \reef{rel1} and \reef{rel2} to find the amplitude \reef{amp} in terms of momenta and polarizations of the external states.

Before performing the traces, we show that the amplitude \reef{amp}   satisfies the Ward identity corresponding to the gauge boson. To this end, we have to  replace the  gauge boson polarization tensor $\z_3^a$ with its momentum $ik_3^a$, and show that  the result vanishes. Using the antisymmetric property of the function $\Gamma^{ab}$, one easily observes that  $a_3,\,a_4,\,a_5,\,a_8,\,a_9,\,a_{10},\,a_{11},\,a_{13}$ become zero after replacing $\z_3^a$ with $ik_3^a$. Using the on-shell relation $p_1\inn k_3=-p_2\inn k_3$, one finds the remaining terms to be 
\beqa
 \cA&\sim&-ib_1 p_1\inn k_3\bigg[- q_1+  q_6- q_{14}\bigg] +ib_2 p_1\inn k_3\bigg[- q_2-q_7+ q_{12}\bigg]
\eeqa
Since $b_1,b_2$ are not zero generally, we should prove that the integrals in each bracket are zero, \ie
\beqa
M_1\equiv-q_1+q_6-q_{14}=\int\frac{K(z_{1\bar1}z_{32}z_{3\bar2}+z_{31}z_{3\bar2}z_{\bar1 2}+z_{31}z_{32}z_{\bar1\bar2})}{z_{12}z_{1\bar{1}}z_{2\bar{2}}z_{31}z_{32}z_{3\bar1}z_{3\bar2}z_{\bar{1}\bar{2}}}d^2z_1d^2z_2dx_3=0\nonumber\\
M_2\equiv-q_2-q_7+q_{12}=\int\frac{K(z_{1\bar1}z_{32}z_{3\bar2}+z_{31}z_{3\bar2}z_{\bar1 2}+z_{31}z_{32}z_{\bar1\bar2})}{z_{12}z_{1\bar{2}}z_{31}z_{32}z_{3\bar1}z_{3\bar2}z_{\bar{1}2}z_{\bar{1}\bar{2}}}d^2z_1d^2z_2dx_3=0\labell{int1}
\eeqa
Using the fact that the  integrands are invariant under the $SL(2,R)$ transformation, one may map the integrands to the   unit disk by the following transformations:
\beqa
z_i\rightarrow -i\frac{z_i-1}{z_i+1}\\
\bar{z_i}\rightarrow i\frac{\bar{z_i}-1}{\bar{z_i}+1}\nonumber
\eeqa
To fix  the $SL(2,R)$ symmetry,    we use the Faddeev-Popove gauge fixing mechanism to fix $z_1=\bar{z_1}=0$ and $x_3=1$. The Jacobian of this transformation is $2i$. Writing   $z_2=re^{i\theta}$ and $\bar{z_2}=re^{-i\theta}$, one finds
\beqa\label{integrals}
M1=\int_0^1rdr\int_0^{2\pi}d\theta\frac{4\sin(\theta)(r^2)^{p_1.p_2}(1-r^2)^{p_2.D.p_2-1}}{r(r^2-2r\cos(\theta)+1)^{2p_1.k_3+1}}\\
M2=\int_0^1rdr\int_0^{2\pi}d\theta\frac{4\sin(\theta)(r^2)^{p_1.p_2}(1-r^2)^{p_2.D.p_2}}{r(r^2-2r\cos(\theta)+1)^{2p_1.k_3+1}}\nonumber
 \eeqa
The $ \theta$ integration then gives zero result. So the amplitude \reef{amp} satisfies the Ward identity corresponding to the gauge boson transformation, as expected.

\subsection{Performing the traces}
We now   calculate the traces in \reef{rel1} and \reef{rel2}. Defining one-traces terms $d_1^{\mu},\,d_3^{\mu b\chi},d_5^{\mu b\chi}$ as:
\beqa
d_1^{\mu}(n)&\equiv&\Tr (P_-\Gamma_{(n)}M_p\gamma^\mu)\nonumber\\
d_3^{\mu b\chi}(n)&\equiv&\Tr(P_-\Gamma_{1(n)}M_p\gamma^\mu\Gamma^{b\chi})\nonumber\\
d_5^{\mu b\chi}(n)&\equiv&\Tr(P_-\Gamma_{1(n)}M_p\Gamma^{b\chi}\gamma^\mu)
\eeqa
the two-trace terms  in \reef{rel1} can be written as 
\beqa
b_1=d_1^{\mu}(n)d_{1\mu}(m),\,b_3^{b\chi}=d_3^{\mu b\chi}(n)d_{1\mu}(m),\,b_5^{b\chi}=d_5^{\mu b\chi}(n)d_{1\mu}(m)\labell{rel3}
\eeqa
The explicit calculation of the traces in $d_1^{\mu},\,d_3^{\mu b\chi},d_3^{\mu b\chi}$ gives the following result:
\beqa
d_1^{\mu}(n)&=&\frac{16}{n!}(-1)^{\frac{1}{2}n(n+1)}\epsilon_{a_0\cdots a_p}\bigg[\delta_{np}\delta^{a_0}_\mu F^{a_1\cdots a_p}+n\delta_{n,p+2} F^{a_0\cdots a_p} {}{} ^{\mu}\bigg]+\bigg(n\rightarrow n', F\rightarrow *F\bigg)\nonumber\\
 d_3^{\mu b\chi}(n)&=&-(-1)^{\frac{1}{2}n(n-1)}\frac{16}{n!}\bigg\{\delta_{n,p-2}F_{a_0 a_1 \cdots a_{p-3}}\veps^{a_0 \cdots a_{p-3} \mu b \chi}+\delta_{np}\bigg[n F_{a_0 a_1 \cdots a_{p-2}}{}{}^{\chi}\veps^{a_0 \cdots a_{p-2} b \mu}\nonumber\\
&&-n F_{a_0 \cdots a_{p-2}}^{}{}^{\mu} \veps^{a_0 a_1 a_2\cdots b \chi}+F_{a_0 \cdots a_{p-1}}\veps^{a_0 \cdots a_{p-1} \chi}\delta^{b\mu}-n F_{a_0 \cdots a_{p-2}}^{}{}^{}{}^{b}\veps^{a_0 \cdots a_{p-2} \chi \mu}\nonumber\\
&&- F_{a_1 \cdots a_p}\veps^{a_1\cdots a_p b}\delta^{\chi \mu}\bigg]-\delta_{n,p+2}n\bigg[(n-1) \veps^{\chi}{}{}_{a_0 a_1\cdots a_{p-1}}F^{b a_0 \cdots a_{p-1} \mu}\nonumber\\
&&-(n-1) \veps_{a_0 \cdots a_{p-1}}^{}{}^{\mu} F^{a_0 \cdots a_{p-1} b\chi}-\veps_{a_0 \cdots a_{p}}F^{a_0 \cdots a_{p}\chi}\delta^{b\mu}\labell{ds}\\
&&-(n-1) \veps_{a_0\cdots a_{p-1}}^{}{}^{}{}^{b}F^{a_0 \cdots a_{p-1}\chi\mu}+ \veps_{a_0\cdots a_p}F^{a_0\cdots a_p b}\delta^{\chi \mu}\bigg]\bigg\}+\bigg(n\rightarrow n', F\rightarrow*F\bigg)\nonumber\\
 d_5^{\mu b\chi}(n)&=&-(-1)^{\frac{1}{2}n(n-1)}\frac{16}{n!}\bigg\{\delta_{n,p-2}F_{a_0 a_1\cdots a_{p-3}}\veps^{a_0 \cdots a_{p-3} \mu b\chi}+\delta_{n,p}\bigg[n F_{a_0 a_1 \cdots a_{p-2}}^{}{}^{\chi}\veps^{a_0 \cdots a_{p-2} b \mu}\nonumber\\
&&-n F_{a_0 \cdots a_{p-2}}^{}{}^{\mu} \veps^{a_0 a_1 a_2\cdots b \chi}-F_{a_0\cdots a_{p-1}}\veps^{a_0\cdots a_{p-1}\chi}\delta^{b\mu}-n F_{a_0\cdots a_{p-2}}^{}{}^{}{}^{b}\veps^{a_0 \cdots a_{p-2} \chi \mu}\nonumber\\
&&+ F_{a_1 \cdots a_p}\veps^{a_1 \cdots a_p b}\delta^{\chi \mu}\bigg]-n\delta_{n,p+2}\bigg[(n-1) \veps_{a_0 a_1 \cdots a_{p-1}}{}{}^{\chi}F^{a_0\cdots a_{p-1} b \mu}\nonumber\\
&&-(n-1) \veps_{a_0\cdots a_{p-1}}^{}{}^{\mu} F^{a_0\cdots a_{p-1} b\chi}+\veps_{a_0 \cdots a_{p}}F^{a_0\cdots a_{p}\chi}\delta^{b\mu}\nonumber\\
&&-(n-1) \veps_{a_0\cdots a_{p-1}}^{}{}^{}{}^{b}F^{a_0\cdots a_{p-1} \chi\mu}- \veps_{a_0\cdots a_p}F^{a_0\cdots a_p b}\delta^{\chi \mu}\bigg]\bigg\}+\bigg(n\rightarrow n',F\rightarrow*F\bigg)\nonumber
\eeqa
where $n'=10-n$ and $F$ is the R-R field strength. When we replace them in \reef{rel3}, one of the R-R field strength is $F_1$ and the other one is $F_2$. Replacing the above one-trace terms in \reef{rel3}, one finds  $b_1 ,\,b_3^{  b\chi},b_3^{  b\chi}$ in terms of momenta and polarizations of the external states. The replacement of the resulting $b_1 ,\,b_3^{  b\chi},b_3^{  b\chi}$ in \reef{rel4}, one then finds $a_1,\,a_3,\,a_4,\,a_5,\,a_6,\,a_8,\,a_{14}$.

The explicit calculation of the one-trace terms $c_2 ,\,c_9^{ b\chi},c_{13}^{ b\chi}$ gives the following result:
\beqa
c_2&\!\!\!\!\!=\!\!\!\!\!&\frac{16}{n!}\delta_{n,m}\bigg[
\Tr(D)F^{(n)}_1.F^{(n)}_2-2nD^\lambda{}{}_\rho F_{1\lambda\mu_2\cdots\mu_n}F_2^{\rho\mu_2\cdots\mu_n}\bigg]+\bigg(n\rightarrow n',F_{1}\rightarrow *F_1\bigg)\nonumber\\
 c_9^{b\chi}&\!\!\!\!\!=\!\!\!\!\!&\frac{16}{n!}\delta_{m,n+2}\bigg[\Tr(D)F_{1(n)}^{\mu_1\cdots\mu_n}
F_{2(m)}^{b\chi}{}{}_{\mu_1\cdots\mu_n}+2D^{b\mu} F_{1}^{\mu_1\cdots\mu_n}
F_{2}{}^{\chi}{}{}_{\mu\mu_1\cdots\mu_n}\nonumber\\
&&-2D^{\mu\chi} F_{1}^{\mu_1\cdots\mu_n}
F_{2}^b{}{}_{\mu,\mu_1,\cdots,\mu_n} -2nD_{\mu}^{}{}^{\nu} F_{1}^{\mu,\mu_2,\cdots,\mu_n}
F_{2}{}_{\nu,a,\chi,\mu_2,\cdots,\mu_n}\bigg]\nonumber\\
&&+\frac{16}{(n-1)!}\delta_{mn}\bigg[\Tr(D)F_{1}^b{}{}_{\mu_2\cdots\mu_n}F_{2}^{\chi\mu_2\cdots\mu_n}-\Tr(D)F_{1}^{\chi\mu_2\cdots,\mu_n}F_{2}^b{}{}_{\mu_2\cdots\mu_n}\nonumber\\
&&-2 D^{b\mu}F_{1}{}{}_{\mu \mu_2\cdots\mu_n}F_{2}^{\chi\mu_2\cdots\mu_n}+2D^{\chi}_{}{}_{\mu}F_{1}^{\mu\mu_2\cdots\mu_n}F_{2}^b{}{}_{\mu_2\cdots\mu_n} \nonumber\\
&&+2(n-1)D^\mu{}{}_\nu\bigg( F_{1}^b{}{}_{\mu\mu_3\cdots\mu_n}F_{2}^{\chi\nu\mu_3\cdots\mu_n}-F_{1}^{\chi\nu\mu_2\cdots\mu_n}F_{2}^b{}{}_{\mu\mu_3\cdots\mu_n}\bigg)\bigg]+\bigg(n\rightarrow n',F_1\rightarrow*F_1\bigg)\nonumber\\
c_{13}^{b\chi}&\!\!\!\!\!=\!\!\!\!\!&\frac{16}{n!}\delta_{m,n+2}\bigg[\Tr(D)F_{1}^{\mu_1\cdots\mu_n}
F_{2}^{a\chi}{}{}_{\mu_1\cdots\mu_n}-2D^{b\mu} F_{1}^{\mu_1\cdots\mu_n}
F_{2}^{\chi}{}{}_{\mu\mu_1\cdots\mu_n}\nonumber\\
&&+2D^{\mu\chi} F_{1}^{\mu_1\cdots\mu_n}
F_{2}^b{}{}_{\mu\mu_1\cdots\mu_n} -2nD_{\mu}^{}{}^{\nu}
 F_{1}^{\mu\mu_2\cdots\mu_n}
F_{2}^{b\chi}{}{}_{\nu\mu_2\cdots\mu_n}\bigg]\nonumber\\
&&+\frac{16}{(n-1)!}\delta_{mn}\bigg[\Tr(D)F_{1}^b{}{}_{\mu_2\cdots\mu_n}F_{2}^{\chi\mu_2\cdots\mu_n}-\Tr(D)
F_{1}^{\chi\mu_2\cdots\mu_n}F_{2}^b{}{}_{\mu_2\cdots\mu_n}\nonumber\\
&&+2D^{b\mu}F_{1}{}{}_{\mu\mu_2\cdots\mu_n}F_{2}^{\chi\mu_2\cdots\mu_n}-2D^{\chi}_{}{}_{\mu}F_{1}^{\mu\mu_2\cdots\mu_n}F_{2}^b{}{}_{\mu_2\cdots\mu_n}\labell{cs}\\
&&+2(n-1)D^\mu{}{}_\nu\bigg( F_{1}^b{}{}_{\mu\mu_3\cdots\mu_n}F_{2}^{\chi\nu\mu_3\cdots\mu_n}-F_{1}^{\chi\nu\mu_3\cdots\mu_n}F_{2}^b{}{}_{\mu\mu_3\cdots\mu_n}\bigg)\bigg]+\bigg(n\rightarrow n',F_1\rightarrow*F_1\bigg)\nonumber
\eeqa
Replacing  the above results for  $c_2 ,\,c_9^{ b\chi},c_{13}^{ b\chi}$ and $b_1 ,\,b_3^{  b\chi},b_5^{  b\chi}$ in \reef{rel}, one finds  $b_2 ,\,b_9^{ b\chi},b_{13}^{ b\chi}$ and hence one finds     $a_2,\,a_7,\,a_9,\,a_{10},\,a_{11},\,a_{12},\,a_{13}$ upon replacing them in \reef{rel5}. Having performed the traces in \reef{amp}, one now has  the amplitude \reef{amp} in terms of momenta and polarization of the external states. To check our results, in the next section, we study the S-duality Ward identity of the amplitude.

\section{S-duality Ward identity}

The D$_p$-brane S-matrix element in the previous section is valid for any $p$. It is known that D$_3$-brane is invariant under S-duality transformation, so the S-matrix elements of D$_3$-brane should satisfy the Ward identity corresponding to the $SL(2,R)$ transformation which is linear transformation on quantum fluctuations and nonlinear transformation on the background fields \cite{Garousi:2012gh}. The B-field and RR two-form transforms as doublet, \ie
\beqa
\cB&\equiv&\pmatrix{B \cr 
C^{(2)}}\rightarrow (\Lambda^{-1})^T \pmatrix{B \cr 
C^{(2)}}\,\,\,;\,\,\,\Lambda=\pmatrix{p&q \cr 
r&s}\in SL(2,R)\labell{2}
\eeqa
The gauge boson field strength $F_{ab}$ and its magnetic dual $(*F)_{ab}=\eps_{abcd}F^{cd}/2$ also transform as doublet, \ie 
\beqa
\cF&\equiv&\pmatrix{(*f) \cr 
e^{-\phi_0}f-C_0(*f)}\rightarrow (\Lambda^{-1})^T \pmatrix{(*f) \cr 
e^{-\phi_0}f-C_0(*f)}\nonumber
\eeqa
where $\phi_0$ and $C_0$ are the background   dilaton and R-R scalar, respectively. The transformation of the background dilaton and R-R scalar is \cite{Gibbons:1995ap,Tseytlin:1996it,Green:1996qg}
\beqa
{\cal M}_0\rightarrow \Lambda {\cal M}_0\Lambda ^T
\eeqa
where the matrix $\cM_0$ is 
\beqa
 {\cal M}_0=e^{\phi_0}\pmatrix{|\tau_0|^2&C_0 \cr 
C_0&1}\labell{M}
\eeqa
where   $\tau_0=C_0+ie^{-\phi_0}$. Quantum fluctuations of the dilaton and the R-R scalar appear in $\delta\cM$, \ie
\beqa
\delta\cM=\pmatrix{-(e^{-\phi_0}-C_0^2e^{\phi_0})\delta\phi+2C_0 e^{\phi_0}\delta C& C_0 e^{\phi_0}\delta\phi+e^{\phi_0}\delta C\cr 
C_0 e^{\phi_0}\delta\phi+e^{\phi_0}\delta C&e^{\phi_0}\delta \phi}\labell{dM}
\eeqa
It also transforms as 
\beqa
{\delta\cal M}\rightarrow \Lambda \delta{\cal M}\Lambda ^T\labell{rel7}
\eeqa
The transverse scalar fields, graviton in the Einstein frame and the R-R four-form are invariant under the S-duality.

Using the above transformations, one should be able to construct a set of  S-matrix elements in terms of product of the above $SL(2,R)$ tensors  such that they make an  invariant under the $SL(2,R)$ transformations. For example, $(*\cF)^T\delta\cM\cB$ is invariant under the linear $SL(2,R)$ transformations. It has the following six   elements \cite{Garousi:2012gh}:
\beqa
(*\cF)^T\delta\cM\cB&=&e^{-\phi_0}\delta\phi fB+\delta\phi(*f)C^{(2)}+C_0\delta\phi(*f)B\nonumber\\
&&+\delta C(*f)B-e^{\phi_0}C_0\delta CfB-e^{\phi_0}\delta C fC^{(2)}\labell{dual}
\eeqa
As a result, the S-matrix of the above six terms should have the same structure.  For flat spacetime with no R-R background field, the above S-dual multiplet simplifies to 
\beqa
(*\cF)^T\delta\cM\cB&=&e^{-\phi_0}\delta\phi fB+\delta\phi(*f)C^{(2)} +\delta C(*f)B -e^{\phi_0}\delta C fC^{(2)}\labell{dual2}
\eeqa
By explicit calculation, it has been shown in \cite{Garousi:2012gh} that the S-matrix elements of first three terms in above multiplet have identical structure. We shall show that the S-matrix element of the last term also has the same structure.   
On the other hand, if an S-matrix element could not be combined with some other S-matrix element to be written in terms of $SL(2,R)$ invariant, that S-matrix element should be zero. In the following subsections,   we fix $p=3$ and examine S-duality Ward identity of the amplitude \reef{amp} for various R-R fields.

\subsection{$C^{(0)}C^{(2)}f$}

 When  $n=1$ and $m=3$ the trace parts of  $a_1, a_2, a_4, a_6, a_7, a_8, a_{12}$ and $a_{14}$ are  zero. The non-zero  terms yield:
\beqa
 a_3&=& 128C^{(0)} p_1.V.F^{(3)}_{ba}f^{ab} \nonumber\\
 a_5&=&- 128C^{(0)} p_1.V.F^{(3)}_{ba}f^{ab} \nonumber\\
 a_9&=& -64C^{(0)} p_1.V.F^{(3)}_{ba}f^{ab}+32C^{(0)} p_1.N.F^{(3)}_{ba}f^{ab} \nonumber\\
 a_{10}&=& 32C^{(0)} p_1.V.F^{(3)}_{ba}f^{ab} \nonumber\\
 a_{11}&=&  64C^{(0)} p_1.V.F^{(3)}_{ba}f^{ab}-32C^{(0)} p_1.N.F^{(3)}_{ba}f^{ab} \nonumber\\
a_{13}&=& 32C^{(0)} p_1.V.F^{(3)}_{ba}f^{ab} 
\eeqa
where $f^{ab}=i(k_3^a\z_3^b-k_3^b\z_3^a)$ is the gauge boson field strength in momentum space, $F^{(3)}_{\mu\nu\alpha}$ is the R-R two-form field strength and $C^{(0}$ is the polarization of the R-R scalar. The matrix $V$ is the world volume metric and $N$ is the transverse space metric, \ie $\eta_{\mu\nu}=V_{\mu\nu}+N_{\mu\nu}, D_{\mu\nu}=V_{\mu\nu}-N_{\mu\nu}$. Replacing them in \reef{amp}, one finds
\beqa
\cA_{C^{(0)}C^{(2)}F}&\sim&\int d^2z_1d^2z_2 dz_3  e^{-3\phi_0/2}C^{(0)}Kf^{ab}\bigg[p_1.N.F^{(3)}_{ba}\frac{z_{\bar{1}1}}{z_{12}z_{13}z_{2\bar{1}}z_{3\bar{1}}z_{1\bar{2}}z_{\bar{1}\bar{2}}} +p_1.V.F^{(3)}_{ba} \labell{eq13}\\
&\times&\frac{(-2 z_3 z_{\bar{1}} + z_2 (z_{3\bar{2}} + z_{\bar{1}\bar{2}} ) + (z_3 + z_{\bar{1}}) z_{\bar{2}}) (z_2 (z_3 - 2 z_{\bar{2}}) + 
   z_3 z_{\bar{2}} + z_1 (z_{23} + z_{\bar{2}3}))}{z_{12}z_{13}z_{23}z_{2\bar{1}}
   z_{3\bar{1}}z_{1\bar{2}}z_{2\bar{2}}z_{3\bar{2}}z_{\bar{1}\bar{2}}}\bigg]\nonumber
\eeqa
where we have also transformed the amplitude to the Einstein frame. This amplitude should have the same structure as the amplitude of first three terms in \reef{dual2}. 

The amplitude of one dilaton, one B-field and one gauge boson in the Einstein frame has been calculated in \cite{Garousi:2012gh} to be 
\beqa
A_{\phi BF}&\sim& \int d^2z_1d^2z_2dz_3\phi_1 e^{-3\phi_0/2}f^{ab}\left(I_{11}\bigg[\frac{p_1.D.p_1}{p_1.p_2}  p_1.V.H_{ba}+\frac{p_1.k_3\,p_1.D.p_1}{(p_1.p_2)^2}p_1.H_{ba}\bigg]\right.\nonumber\\
 &&\left.-I_2\bigg[4p_1.V.H_{ba}+\frac{p_2.D.p_2}{p_1.p_2} \left(2 p_1.H_{ba}- p_1.N.H_{ba}\right)+\frac{p_1.k_3\,p_2.D.p_2}{(p_1.p_2)^2}p_1.H_{ba}\bigg]\right) \labell{amp32}
\eeqa
where $\phi_1$ is the polarization of the dilaton, $H$ is the B-field strength and 
\beqa
I_2&=&\frac{K}{z_{31} z_{2 \bar{1}} z_{3 \bar{1}} z_{1\bar{2}} z_{2 \bar{2}}}\nonumber\\
I_{11}&=&\frac{K}{z_{32}z_{1\bar{1}} z_{2 \bar{1}} z_{1\bar{2}} z_{3 \bar{2}}}\labell{int.1} 
\eeqa
 These integrals satisfy the following two relations:
\beqa
&&4 I_2 k_3.p_1+\left(2 I_2+I_3\right) p_1.p_2+\left(I_2-I_{11}\right) p_1.D.p_1=0,\nonumber\\
&&2 I_3 k_3.p_1+\left(-2 I_1+I_3\right) p_1.p_2-\left(I_2+I_{11}\right) p_1.D.p_1=0\labell{rel.2}
\eeqa
where
\beqa
I_{1}&=&\frac{K}{z_{12} z_{31} z_{3 \bar{1}} z_{2 \bar{2}} z_{\bar{1} \bar{2}}}\labell{int.2}\\
I_3&=&\frac{K}{z_{12} z_{32} z_{1\bar{1}} z_{3 \bar{1}} z_{1\bar{2}}}-\frac{2 K}{z_{12} z_{31} z_{2 \bar{1}} z_{3 \bar{1}} z_{2 \bar{2}}}+\frac{K}{z_{12} z_{31} z_{2 \bar{1}} z_{3 \bar{1}} z_{3 \bar{2}}}+\frac{K}{z_{31} z_{32} z_{1\bar{1}} z_{2 \bar{1}} z_{\bar{1} \bar{2}}}\nonumber
\eeqa
To compare the amplitude \reef{amp32} with \reef{eq13}, we solve the above two relations to find 
\beqa
\frac{p_1.D.p_1}{p_1.p_2}&=&\frac{\left(2 I_2+I_3\right) \times 2I_3-\left(-2 I_1+I_3\right)\times 4I_2}{-\left(I_2+I_{11}\right)\times 4I_2-\left(I_2-I_{11}\right) \times 2I_3}\nonumber\\
 \frac{p_1.k_3}{p_1.p_2}&=&\frac{\left(-2 I_1+I_3\right)\left(I_2-I_{11}\right)+\left(2 I_2+I_3\right)\left(I_2+I_{11}\right)}{-\left(I_2+I_{11}\right)\times 4I_2-\left(I_2-I_{11}\right) \times 2I_3}\labell{rel6}
\eeqa
Now using the on-shell relation
\beqa
p_2.D.p_2=p_1.D.p_1+4p_1.k_3\labell{rel71}
\eeqa
to write $p_2.D.p_2$ in \reef{amp32} in terms of $p_1.D.p_1$ and $p_1.k_3$, and then using the relations in \reef{rel6}, one can simplify the amplitude \reef{amp32} as
\beqa
A_{\phi BF}&\sim& \int d^2z_1d^2z_2dz_3\, \phi_1e^{-3\phi_0/2}f^{ab}\bigg[p_1.V.H_{ba} 
(I_1-I_2+I_3)+
p_1.N.H_{ba}( I_1+I_2)\bigg]\nonumber 
\eeqa
Using  \reef{int.1} and \reef{int.2},  one finds  exactly the structure in \reef{eq13}. The extra factor of $e^{2\phi_0}$ in the last term in \reef{dual2} compared to the first term, is that in the study of the S-duality the  R-R fields should be rescaled as $C\rightarrow e^{\phi}C$.

\subsection{$C^{(0)}C^{(0)}f$}

Since the R-R scalar transforms as \reef{rel7} and the gauge boson field strength transforms as doublet, it is impossible to construct an $SL(2,R)$ invariant combination from two $\delta \cM$ and one gauge boson. As a result, the amplitude of two R-R scalars and one gauge boson must be zero. Using the identity \reef{int1}, one finds  the contribution of $A_1,A_2,A_6,A_7,A_{12}$ and $A_{14}$ to \reef{amp0} is zero. The trace $d_1^{\mu}$ is zero, so $b_1,b_3^{b\chi},b_5^{b\chi}$ are zero. So the contributions of $A_1,A_3,A_4,A_6,A_{8}$ and $A_{14}$ are zero. Moreover, using the identity \reef{int1}, one finds  the contributions of $A_2,A_7,A_{12}$ to \reef{amp0} are also zero. The traces in the remaining four terms produce $p_1^a p_2^b f_{ab}$ which is zero using   momentum conservation $p_2^a=-(p_1^a+k^a)$ and the on-shell relation $k_3^af_{ab}=0$.

\subsection{$C^{(2)}C^{(2)}f$}
One can not construct an $SL(2,R)$ invariant from three doublets, so the amplitude of two R-R two-form and one gauge boson must be zero. Using the identity \reef{int1}, one finds zero contribution from  $ A_1,A_2,A_6,A_7,A_{12}$ and $A_{14}$. The    nonzero terms are:
\beqa
A_3&=& 128 f^{ab}F_1^{(3)}{}{}_{acd}F_2^{(3)}{}{}_b{}{}^{cd} q_3\\
A_4&=&- 128 f^{ab}F_1^{(3)}{}{}_{acd}F_2^{(3)}{}{}_b{}{}^{cd}q_4\nonumber\\
A_5&=& 128 f^{ab}F_1^{(3)}{}{}_{acd}F_2^{(3)}{}{}_b{}{}^{cd}q_5\nonumber\\
A_8&=&- 128 f^{ab}F_1^{(3)}{}{}_{acd}F_2^{(3)}{}{}_b{}{}^{cd}q_8\nonumber\\
A_9&=& 64 f^{ab}F_1^{(3)}{}{}_{ac\mu}F_2^{(3)}{}{}_b{}{}^{c\mu}q_9\nonumber\\
A_{10}&=&- 64 f^{ab}F_1^{(3)}{}{}_{ac\mu}F_2^{(3)}{}{}_b{}{}^{c\mu}q_{10} \nonumber\\
A_{11}&=& 64 f^{ab}F_1^{(3)}{}{}_{ac\mu}F_2^{(3)}{}{}_b{}{}^{c\mu}q_{11}\nonumber\\
A_{13}&=&- 64 f^{ab}F_1^{(3)}{}{}_{ac\mu}F_2^{(3)}{}{}_b{}{}^{c\mu}q_{13}\nonumber
\eeqa
The amplitude \reef{amp} simplifies to
\beqa
\cA&\sim&128f^{ab}F_1^{(3)}{}{}_{acd}F_2^{(3)}{}{}_b{}{}^{cd} M_3+64f^{ab}F_1^{(3)}{}{}_{ac\mu}F_2^{(3)}{}{}_b{}{}^{c\mu}M_4
\eeqa
where
\beqa
M_3&\equiv&q_3-q_4+q_5-q_8=\int\frac {K[z_ {32} z_ {3\bar1} z_ {3\bar2} - z_ {31} (z_ {32}(z_ {3\bar1} - z_ {3\bar2}) + 
z_ {3\bar1} z_ {3\bar2})]} {z_ {12} z_ {1\bar1} z_ {2\bar2} z_{31} z_ {32} z_{3\bar1} z_ {3\bar2} z_ {\bar1\bar2}}d^2z_1d^2z_2dx_3\labell{int2}\\
M_4&\equiv&q_9-q_{10}+q_{11}-q_{13}=\int\frac {K[z_ {32} z_ {3\bar1} z_ {3\bar2} - z_ {31} (z_ {32}(z_ {3\bar1} - z_{3\bar2}) +z_ {3\bar1} z_ {3\bar2})]} {z_ {12} z_ {1\bar2}z_ {31} z_ {32} z_{3\bar1} z_ {3\bar2} z_ {\bar12} z_{\bar1\bar2}}d^2z_1d^2z_2dx_3\nonumber
\eeqa
Using the same step that we have done for $M_1$ and $M_2$ integrals, one finds
\beqa
M_3=\int_0^1rdr\int_0^{2\pi}d\theta\frac{2\sin(\theta)(r^2)^{p_1.p_2}(1-r^2)^{p_2.D.p_2-1}}{r(r^2-2r\cos(\theta)+1)^{2p_1.k_3+1}}\\
M_4=\int_0^1rdr\int_0^{2\pi}d\theta\frac{2\sin(\theta)(r^2)^{p_1.p_2}(1-r^2)^{p_2.D.p_2}}{r(r^2-2r\cos(\theta)+1)^{2p_1.k_3+1}}\nonumber
 \eeqa
The $\theta$ integration again  gives zero result. Therefore, as the gauge  symmetry Ward identity predicts the constrain \reef{int1} between the $q_i$'s, the above S-duality Ward identity produces the following constrains:
\beqa\label{cons2}
q_3-q_4+q_5-q_8=0\nonumber\\
q_9-q_{10}+q_{11}-q_{13}=0\labell{int3}
\eeqa
One may use the above constrains and the constrains in \reef{int1} to simplify the amplitude \reef{amp}.

\subsection{$C^{(4)}C^{(4)}f$}

The R-R four-form is invariant under the S-duality and $f$ transforms as doublet, so the amplitude of two R-R four-forms and one gauge field can not combined with any S-matrix element to be invariant under the $SL(2,R)$ transformation. As a result, this amplitude must be zero for D$_3$-brane. The explicit calculations produce the following non-zero terms for $A_i$'s: 
\beqa
A_3&=& \frac{128}{3} f^{ab}F_1^{(5)}{}{}_{acde\mu}F_2^{(5)}{}{}_b{}{}^{cde\mu}q_3\\
A_4&=& -\frac{128}{3} f^{ab}F_1^{(5)}{}{}_{acde\mu}F_2^{(5)}{}{}_b{}{}^{cde\mu}q_4\nonumber\\
A_5&=& \frac{128}{3}f^{ab}F_1^{(5)}{}{}_{acde\mu}F_2^{(5)}{}{}_b{}{}^{cde\mu}q_5\nonumber\\
A_8&=& -\frac{128}{3} f^{ab}F_1^{(5)}{}{}_{acde\mu}F_2^{(5)}{}{}_b{}{}^{cde\mu}q_8\nonumber\\
A_9&=&- \frac{16}{6}f^{ab}\bigg[F_1^{(5)}{}{}_{a\mu\nu\rho\sigma}F_2^{(5)}{}{}_b{}{}^{\mu\nu\rho\sigma}+2D^{\mu\nu}F_1^{(5)}{}{}_{a\mu\rho\sigma\delta}F_2^{(5)}{}{}_{b\nu}{}{}^{\rho\sigma\delta}\bigg]q_9\nonumber\\
A_{10}&=& \frac{16}{6}f^{ab}\bigg[F_1^{(5)}{}{}_{a\mu\nu\rho\sigma}F_2^{(5)}{}{}_b{}{}^{\mu\nu\rho\sigma}+2D^{\mu\nu}F_1^{(5)}{}{}_{a\mu\rho\sigma\delta}F_2^{(5)}{}{}_{b\nu}{}{}^{\rho\sigma\delta}\bigg]q_{10} \nonumber\\
A_{11}&=&- \frac{16}{6}f^{ab}\bigg[F_1^{(5)}{}{}_{a\mu\nu\rho\sigma}F_2^{(5)}{}{}_b{}{}^{\mu\nu\rho\sigma}+2D^{\mu\nu}F_1^{(5)}{}{}_{a\mu\rho\sigma\delta}F_2^{(5)}{}{}_{b\nu}{}{}^{\rho\sigma\delta}\bigg]q_{11}\nonumber\\
A_{13}&=&\frac{16}{6}f^{ab}\bigg[F_1^{(5)}{}{}_{a\mu\nu\rho\sigma}F_2^{(5)}{}{}_b{}{}^{\mu\nu\rho\sigma}+2D^{\mu\nu}F_1^{(5)}{}{}_{a\mu\rho\sigma\delta}F_2^{(5)}{}{}_{b\nu}{}{}^{\rho\sigma\delta}\bigg]q_{13}\nonumber
\eeqa
Replacing them in \reef{amp}, one finds:
\beqa
\cA \sim\frac{128}{3}f^{ab}F_1^{(5)}{}{}_{acde\mu}F_2^{(5)}{}{}_b{}{}^{cde\mu}M_3 +\frac{16}{6}f^{ab}\bigg[F_1^{(5)}{}{}_{a\mu\nu\rho\sigma}F_2^{(5)}{}{}_b{}{}^{\mu\nu\rho\sigma}+2D^{\mu\nu}F_1^{(5)}{}{}_{a\mu\rho\sigma\delta}F_2^{(5)}{}{}_{b\nu}{}{}^{\rho\sigma\delta}\bigg]M_4\nonumber
\eeqa
where $M_3,\, M_4$ are given in \reef{int2}. Since they are zero, \ie \reef{int3}, the above amplitude is zero, as expected from the S-duality Ward identity.
 
\subsection{$C^{(0)}C^{(2)}\Phi$}

The amplitude of one R-R scalar, one R-R two-form and one transverse scalar must be zero because $C^{(0)}$ transforms as a modular, $C^{(2)}$ transforms as a doublet and $\Phi$ transforms as a scalar under S-duality transformations. The only non-zero $A_i$'s are the following:  
\beqa
A_9&=&- 128iF^{(1)}{}{}^{\mu}F^{(3)}{}{}_{\mu a i}k^a\zeta^iq_9\\
A_{10}&=& 64iF^{(1)}{}{}^{\mu}F^{(3)}{}{}_{\mu a i}k^a\zeta^iq_{10}\nonumber\\
A_{11}&=&- 128iF^{(1)}{}{}^{\mu}F^{(3)}{}{}_{\mu a i}k^a\zeta^iq_{11}\nonumber\\
A_{13}&=& 64iF^{(1)}{}{}^{\mu}F^{(3)}{}{}_{\mu a i}k^a\zeta^iq_{13}\nonumber
\eeqa
The amplitude \reef{amp}, then becomes
\beqa
\cA&\sim&64iF^{(1)}{}{}^{\mu}F^{(3)}{}{}_{\mu a i}k^a\zeta^i M_5
\eeqa
where
\beqa
M_5&\equiv&-2q_9+q_{10}-2q_{11}+q_{13}\\
&&=\int\frac{K[z_{31}z_{32}z_{3\bar1}-2z_{31}z_{32}z_{3\bar2}+
z_{31}z_{3\bar1}z_{3\bar2}-2z_{32}z_{3\bar1}z_{3\bar2}]}{z_{12}z_{1\bar2}z_{31}z_{32}z_{3\bar1}z_{3\bar2}z_{\bar12}z_{\bar1\bar2}}d^2z_1d^2z_2dx_3\nonumber
\eeqa
Transforming it to the unit disk and fixing the position of one closed string state and the open string states, one finds 
\beqa
M_5&=&\int_0^1rdr\int_0^{2\pi}d\theta\frac{4\sin(\theta)(r^2)^{p_1.p_2}(1-r^2)^{p_2.D.p_2}}{r(r^2-2r\cos(\theta)+1)^{2p_1.k_3+1}}
 \eeqa
Again the $\theta$ integration gives zero result. So the above S-duality Ward identity produces the following constraint:
\beqa\label{cons3}
-2q_9+q_{10}-2q_{11}+q_{13}=0\labell{int4}
\eeqa
which is verified explicitly.

\subsection{$C^{(0)}C^{(4)}\Phi$}

Since $C^{(0)}$ transforms as a modular and $C^{(4)}$ and  $\Phi$ transform  as  scalars under the S-duality transformation, the S-duality Ward identity predicts that the amplitude of one R-R scalar, one R-R four-form and one transverse scalar must be zero . In this case, all trace parts of the amplitude are in fact zero.

\subsection{$C^{(2)}C^{(4)}\Phi$}
The S-duality Ward identity also predicts that the amplitude of one R-R scalar, one R-R two-form and one transverse scalar is zero.   $C^{(4)}$ and  $\Phi$ transform  as a scalar but $C^{(2)}$ transform as a doublet under S-duality transformation. The non zero $A_i$'s are the following:
\beqa
A_3&=&-\frac{128i}{3}F^{(3)}{}{}^{bcd}F^{(5)}_{abcdi}k^a\zeta^iq_3\\
A_4&=&-\frac{128i}{3}F^{(3)}{}{}^{bcd}F^{(5)}_{abcdi}k^a\zeta^iq_4\nonumber\\
A_5&=&-\frac{128i}{3}F^{(3)}{}{}^{bcd}F^{(5)}_{abcdi}k^a\zeta^iq_5\nonumber\\
A_8&=&-\frac{128i}{3}F^{(3)}{}{}^{bcd}F^{(5)}_{abcdi}k^a\zeta^iq_8\nonumber\\
A_9&=& ik^a\zeta^i\bigg[32F^{(3)}{}{}^{\mu\nu\rho}F^{(5)}_{\mu\nu\rho a i}+96F^{(3)}{}{}^{b\mu j}F^{(5)}_{b\mu jai}\bigg]q_9\nonumber\\
A_{10}&=&-\frac{2i}{3} k^a\zeta^i\bigg[32F^{(3)}{}{}^{\mu\nu\rho}F^{(5)}_{\mu\nu\rho a i}+96F^{(3)}{}{}^{b\mu j}F^{(5)}_{b\mu jai}\bigg]q_{10} \nonumber\\
A_{11}&=& ik^a\zeta^i\bigg[32F^{(3)}{}{}^{\mu\nu\rho}F^{(5)}_{\mu\nu\rho a i}+96F^{(3)}{}{}^{b\mu j}F^{(5)}_{b\mu jai}\bigg]q_{11}\nonumber\\
A_{13}&=&-\frac{2i}{3}k^a\zeta^i\bigg[32F^{(3)}{}{}^{\mu\nu\rho}F^{(5)}_{\mu\nu\rho a i}+96F^{(3)}{}{}^{b\mu j}F^{(5)}_{b\mu jai}\bigg]q_{13}\nonumber
\eeqa 
Using the constraints \reef{int3} and \reef{int4}, one finds the amplitude simplifies to 
\beqa
\cA&\sim&\frac{128}{3}iF^{(3)}{}{}^{bcd}F^{(5)}_{abcdi}k^a\zeta^i M_6 
\eeqa
where
\beqa
M_9&\equiv&q_3+q_4+q_5+q_8\nonumber\\
&=&\int-\frac{K[z_{31}(z_{32}(z_{3\bar1}+z_{3\bar2})+z_{3\bar1}z_{3\bar2})+z_{32}z_{3\bar1}z_{3\bar2}]}{z_{12}z_{1\bar1}z_{2\bar2}z_{31}z_{32}z_{3\bar1}z_{3\bar2}z_{\bar1\bar2}}
d^2z_1d^2z_2dx_3 \nonumber\\
&\rightarrow& \int_0^1rdr\int_0^{2\pi}d\theta\frac{4\sin(\theta)(r^2)^{p_1.p_2}(1-r^2)^{p_2.D.p_2-1}}{r(r^2-2r\cos(\theta)+1)^{2p_1.k_3+1}} 
\eeqa
which is zero upon integrating over the  $\theta$ variable. So the Ward identity predicts another constraint, \ie 
\beqa\label{cons4}
q_3+q_4+q_5+q_8=0\labell{int5}
\eeqa
which is verified by the explicit calculation.

We have seen that the amplitudes that are constrained by the S-duality Ward identity to be zero, are in fact zero by explicit calculations. However, there are  other amplitudes in D$_3$-brane that the Ward identity does not predict them to be zero. They are either invariant under the $SL(2,R)$ transformations or they are related to the other amplitudes that involve NS-NS closed string states. In the next section we are going to write the explicit form of these amplitudes and the amplitudes for $p\neq 3$. 
 
\section{Non-zero amplitudes}

The Ward identities predict the six constraints \reef{int1}, \reef{int3}, \reef{int4} and \reef{int5} between the integrals that appear in \reef{amp}. Examining the integrals, we have also found the following four relations between them:
\beqa
q_1+q_5-q_8&=&0\\
q_4-q_5-q_{14}&=&0\nonumber\\
q_6+2q_5&=&0\nonumber\\
q_{12}+2q_{11}&=&0\nonumber
\eeqa
Using these ten constraints, one  can express all integrals in \reef{amp} in terms of the following four  integrals:
\beqa
Q_1&\equiv&q_6\rightarrow\int_0^12rdr\int_0^{2\pi}d\theta\frac{(r^2)^{p_1.p_2-1}(1-r^2)^{p_2.D.p_2-1}}{(r^2-2r\cos(\theta)+1)^{2p_1.k_3}}\nonumber\\
Q_2&\equiv&q_{14}-q_1\rightarrow\int_0^12rdr\int_0^{2\pi}d\theta\frac{(r^2)^{p_1.p_2-1}(1-r^2)^{p_2.D.p_2}}{(r^2-2r\cos(\theta)+1)^{2p_1.k_3+1}}\nonumber\\
Q_3&\equiv&q_{12}\rightarrow\int_0^12rdr\int_0^{2\pi}d\theta\frac{(r^2)^{p_1.p_2-1}(1-r^2)^{p_2.D.p_2}}{(r^2-2r\cos(\theta)+1)^{2p_1.k_3}}\nonumber\\
Q_4&\equiv&q_7-q_2\rightarrow\int_0^12rdr\int_0^{2\pi}d\theta\frac{(r^2)^{p_1.p_2-1}(1-r^2)^{p_2.D.p_2+1}}{(r^2-2r\cos(\theta)+1)^{2p_1.k_3+1}}\nonumber
\eeqa
 The $\theta$-integral can be evaluated using the following formula \cite{ISG}:
\beqa
\int_0^{2\pi}d\theta\frac{\cos(n\theta)}{(1+x^2-2x\cos(\theta))^b}&=&2\pi x^n\frac{\Ga(b+n)}{n!\Ga(b)}{}_2F_1\bigg[{b, \ n+b \atop n+1}\ ;\ x^2\bigg]
\eeqa
where $|x|<1$. The $r$-integral can be evaluated using the following formula \cite{ISG}:
\beqa
\int_0^1dx\,x^a(1-x)^b{}_{2}F_{1}\bigg[{ a_1, \  a_2\atop   b_1}\ ;\  x
\bigg]&\!\!\!\!\!=\!\!\!\!\!&{}_{3}F_{2}\bigg[{1+a,\ a_1, \  a_2\atop  2+a+b,\ b_1}\ ;\ 1
\bigg]B(1+a,1+b)\nonumber
\eeqa
Then $Q_i$'s   becomes
\beqa
Q_1&=&2\pi B(p_1.p_2,p_2.D.p_2)\,{}_{3}F_{2}\bigg[{p_1.p_2,\ 2p_1.k_3, \  2p_1.k_3\atop  p_1.p_2+p_2.D.p_2,\ 1}\ ;\ 1
\bigg]\nonumber\\
Q_2&=&2\pi B(p_1.p_2,p_2.D.p_2+1)\,{}_{3}F_{2}\bigg[{p_1.p_2,\ 2p_1.k_3+1, \  2p_1.k_3+1\atop  p_1.p_2+p_2.D.p_2+1,\ 1}\ ;\ 1
\bigg]\nonumber\\
Q_3&=&2\pi B(p_1.p_2,p_2.D.p_2+1)\,{}_{3}F_{2}\bigg[{p_1.p_2,\ 2p_1.k_3, \  2p_1.k_3\atop  p_1.p_2+p_2.D.p_2+1,\ 1}\ ;\ 1
\bigg]\nonumber\\
Q_4&=&2\pi B(p_1.p_2,p_2.D.p_2+2)\,{}_{3}F_{2}\bigg[{p_1.p_2,\ 2p_1.k_3+1, \  2p_1.k_3+1\atop  p_1.p_2+p_2.D.p_2+2,\ 1}\ ;\ 1
\bigg]
\eeqa
Using the package \cite{Huber:2005yg}, one may expand the integrals at low energy, \ie
\beqa
&&Q_1=\frac{1}{p_1.p_2}+\frac{1}{p_2.D.p_2}+\frac{\pi^2}{6}\bigg(-p_1.p_2+\frac{4(p_2.k_3)^2}{p_2.D.p_2}-p_2.D.p_2\bigg)+\cdots\nonumber\\
&&Q_2=\frac{1}{p_1.p_2}+\frac{1}{p_1.D.p_1}+\frac{\pi^2}{6}\bigg(-p_1.p_2+\frac{4(p_1.k_3)^2}{p_1.D.p_1}-p_1.D.p_1\bigg)+\cdots\nonumber\\
&&Q_3=\frac{1}{p_1.p_2}+\frac{\pi^2}{6}p_2.D.p_2+\cdots\nonumber\\
&&Q_4=\frac{1}{p_1.p_2}+\frac{\pi^2}{6}p_1.D.p_1+\cdots\nonumber
\eeqa
where we have also used the on-shell relation \reef{rel71}. One may use the above expansions to find four-derivative couplings of two R-R and one NS states in which we are not interested in this paper. By studying the above low energy expansions, we have observed that the integrals $Q_1, Q_2$ interchange under changing the R-R labels, \ie $Q_1(p_1,p_2)\leftrightarrow Q_2(p_2,p_1)$. Similarly, $Q_3(p_1,p_2)\leftrightarrow Q_4(p_2,p_1)$ 

Since the results of the traces in \reef{ds} depend on $p$, it is convenient to write the non-zero amplitudes for explicit $p$. In the next subsections, we write the amplitudes for $p=0,1,2,3$.

\subsection{$p=0$}

When $p=0$, the nonzero amplitudes are:
\beqa\label{c1c1phi0}
\cA_{C^{(1)}C^{(1)}\Phi}&\sim&\frac{256}{1!}\bigg[p_1.N.\zeta (Q_1-Q_3)+p_2.N.\zeta (Q_2-Q_4)\bigg]F_1^{(2)}{}{}^{i a}F_2^{(2)}{}{}_{i a}\nonumber\\
&&+\frac{16}{2!}\bigg[p_1.N.\zeta Q_3+p_2.N.\zeta Q_4\bigg]\bigg(-8F_1^{(2)}{}{}^{\mu \nu}F_2^{(2)}{}{}_{\mu \nu}-4D^\lambda_{}{}_\rho F_1^{(2)}{}{}^{\rho \nu}F_2^{(2)}{}{}_{\lambda \nu}\bigg)\nonumber\\
&&+\bigg[64(4Q_1-2Q_3-Q_4)F_1^{(2)}{}{}^{ij}F_2^{(2)}{}{}_{aj}+64(4Q_2-Q_3-2Q_4)F_1^{(2)}{}{}_{aj}F_2^{(2)}{}{}^{ij}\bigg]k^a \zeta_i\nonumber\\
\cA_{C^{(3)}C^{(3)}\Phi}&\sim&\frac{16}{4!}\bigg[p_1.N.\zeta Q_3+p_2.N.\zeta Q_4\bigg]\bigg(-8F_1^{(4)}{}{}^{\mu \nu \sigma \delta}F_2^{(4)}{}{}_{\mu \nu \sigma \delta}-8D^\lambda_{}{}_\rho F_1^{(4)}{}{}^{\rho \nu \sigma \delta}F_2^{(4)}{}{}_{\lambda \nu \sigma \delta}\bigg)\nonumber\\
&&-\bigg[\frac{32}{3}Q_3 F_1^{(4)}{}{}_i{}{}^{jkl}F_2^{(4)}{}{}_{ajkl}+\frac{32}{3}Q_4 F_1^{(4)}{}{}_{ajkl}F_2^{(4)}{}{}_i{}{}^{jkl}\bigg] k^a \zeta^i\labell{c3c3phi0}
\eeqa
The amplitude are symmetric under interchanging the   particle labels $1,2$, as expected.

There is no non-zero amplitude when the open string is the gauge field which is consistent with the gauge transformation Ward identity because the world-volume indices can take only one value. As a result, $f^{ab}$ is zero.

\subsection{$p=1$}

When $p=1$, the nonzero amplitudes are:
\beqa 
\cA_{C^{(0)}C^{(0)}\Phi}&\!\!\!\!\!\sim\!\!\!\!\!&\frac{256}{1!}\bigg[p_1.N.\zeta (Q_1-Q_3)+p_2.N.\zeta (Q_2-Q_4)\bigg]F_1^{(1)}{}{}^{a}F_2^{(1)}{}{}_{a}\nonumber\\
&&+\frac{16}{1!}\bigg[p_1.N.\zeta Q_3+p_2.N.\zeta Q_4\bigg]\bigg(-6F_1^{(1)}{}{}^{\mu}F_2^{(1)}{}{}_{\mu}-2D^\lambda_{}{}_\rho F_1^{(1)}{}{}^{\rho}F_2^{(1)}{}{}_{\lambda}\bigg)\nonumber\\
&&+\bigg[64(4Q_1-2Q_3-Q_4)F_1^{(1)}{}{}^{i}F_2^{(1)}{}{}^{a}+64(4Q_2-Q_3-2Q_4)F_1^{(1)}{}{}^{a}F_2^{(1)}{}{}^{i}\bigg]k_a\zeta_i\nonumber\\
 \cA_{C^{(2)}C^{(2)}\Phi}&\!\!\!\!\!\sim\!\!\!\!\!&\frac{256}{2!}\bigg[p_1.N.\zeta  (Q_1-Q_3)+p_2.N.\zeta (Q_2-Q_4)\bigg]F_1^{(3)}{}{}^{i ab}F_2^{(3)}{}{}_{i ab}\nonumber\\
&&+\frac{16}{3!}\bigg[p_1.N.\zeta Q_3+p_2.N.\zeta Q_4\bigg]\bigg(-6F_1^{(3)}{}{}^{\mu\nu\rho}F_2^{(3)}{}{}_{\mu\nu\rho}-6D^\lambda_{}{}_\rho F_1^{(3)}{}{}^{\rho\mu\nu}F_2^{(3)}{}{}_{\lambda\mu\nu}\bigg)\nonumber\\
&&+\bigg[64(4Q_1-2Q_3-Q_4)F_1^{(3)}{}{}^{bij}F_2^{(3)}{}{}_{abj}+64(4Q_2-Q_3-2Q_4)F_1^{(3)}{}{}_{abj}F_2^{(3)}{}{}^{bij}\bigg]k^a\zeta_i\nonumber\\
&&-\bigg[32Q_3F_1^{(3)}{}{}^{ijk}F_2^{(3)}{}{}_{ajk}+32Q_4F_1^{(3)}{}{}_{ajk}F_2^{(3)}{}{}^{ijk}\bigg]k^a\zeta_i\nonumber\\
\cA_{C^{(4)}C^{(4)}\Phi}&\!\!\!\!\!\sim\!\!\!\!\!& -\frac{16}{5!}\bigg[p_1.N.\zeta Q_3+p_2.N.\zeta Q_4\bigg]\bigg(6F_1^{(5)}{}{}^{\mu\nu\rho\sigma\delta}F_2^{(5)}{}{}_{\mu\nu\rho\sigma\delta}+10D^\lambda_{}{}_\rho F_1^{(5)}{}{}^{\rho\mu\nu\sigma\delta}F_2^{(5)}{}{}_{\lambda\mu\nu\sigma\delta}\bigg)\nonumber\\
&&-\bigg[\frac{32}{3}Q_3F_1^{(5)}{}{}^{b}{}{}_i{}{}^{jkl}F_2^{(5)}{}{}_{abjkl}+\frac{32}{3}Q_4F_1^{(5)}{}{}_a{}{}^{bjkl}F_2^{(5)}{}{}_{bijkl}\bigg]k^a\zeta^i\nonumber\\
&&-\bigg[\frac{8}{3}Q_4F_1^{(5)}{}{}_i{}{}^{jklm}F_2^{(5)}{}{}_{ajklm}+\frac{8}{3}Q_3F_1^{(5)}{}{}_a{}{}^{jklm}F_2^{(5)}{}{}_{ijklm}\bigg]k^a\zeta^i\nonumber\\
\cA_{C^{(0)}C^{(2)}f}&\!\!\!\!\!\sim\!\!\!\!\!&-32i(4Q_1-2Q_3-Q_4)F_1^{(1)}{}{}^iF_2^{(3)}{}{}_{iab}f^{ab}\nonumber\\
\cA_{C^{(2)}C^{(4)}f}&\!\!\!\!\!\sim\!\!\!\!\!&\frac{16i}{3}Q_3F_1^{(3)}{}{}_{ijk}F_2^{(5)}{}{}_{ab}{}{}^{ijk}f^{ab}\labell{aaa1}
\eeqa
The first three amplitudes in which the degrees of the R-R fields are identical, are symmetric under interchanging the particle labels $1,2$. The last two amplitudes in which the degrees of the R-R fields are not identical, are not symmetric.  When the particle labels are changed, the amplitudes become
\beqa
\cA_{C^{(2)}C^{(0)}f}& \sim &-32i(4Q_2-2Q_4-Q_3)F_2^{(1)}{}{}^iF_1^{(3)}{}{}_{iab}f^{ab}\nonumber\\
\cA_{C^{(4)}C^{(2)}f}& \sim &\frac{16i}{3}Q_4F_2^{(3)}{}{}_{ijk}F_1^{(5)}{}{}_{ab}{}{}^{ijk}f^{ab}\labell{aaa2}
\eeqa

\subsection{$p=2$}

For the case that $p=2$, the nonzero amplitudes are:
\beqa
\cA_{C^{(1)}C^{(1)}\Phi}&\!\!\!\!\!\sim\!\!\!\!\!&\frac{256}{2!}\bigg[p_1.N.\zeta (Q_1-Q_3)+p_2.N.\zeta  (Q_2-Q_4)\bigg]F_1^{(2)}{}{}^{ab}F_2^{(2)}{}{}_{ab}\nonumber\\
&&+\frac{16}{2!}\bigg[p_1.N.\zeta Q_3+p_2.N.\zeta Q_4\bigg]\bigg(-4F_1^{(2)}{}{}^{\mu \nu}F_2^{(2)}{}{}_{\mu \nu}-4D^\lambda_{}{}_\rho F_1^{(2)}{}{}^{\rho \nu}F_2^{(2)}{}{}_{\lambda \nu}\bigg)\nonumber\\
&&+\bigg[64(4Q_1-2Q_3-Q_4)F_1^{(2)}{}{}^{ib}F_2^{(2)}{}{}_{ab}+1\leftrightarrow 2\bigg]k^a\zeta_i\nonumber\\
&&-\bigg[32Q_3F_1^{(2)}{}{}^{ij}F_2^{(2)}{}{}_{aj}+32Q_4F_1^{(2)}{}{}_{aj}F_2^{(2)}{}{}^{ij}\bigg]k^a\zeta_i\nonumber\\
\cA_{C^{(3)}C^{(3)}\Phi}&\!\!\!\!\!\sim\!\!\!\!\!&\frac{256}{3!}\bigg[p_1.N.\zeta (Q_1-Q_3)+p_2.N.\zeta (Q_2-Q_4)\bigg]F_1^{(4)}{}{}^{i abc}F_2^{(4)}{}{}_{i abc}\nonumber\\
&&-\frac{64}{4!}\bigg[p_1.N.\zeta Q_3+p_2.N.\zeta Q_4\bigg]\bigg(F_1^{(4)}{}{}^{\mu \nu \sigma \delta}F_2^{(4)}{}{}_{\mu \nu \sigma \delta}+2D^\lambda_{}{}_\rho F_1^{(4)}{}{}^{\rho \nu \sigma \delta}F_2^{(4)}{}{}_{\lambda \nu \sigma \delta}\bigg)\nonumber\\
&&+\bigg[32(4Q_1-2Q_3-Q_4)F_1^{(4)}{}{}^{bc}{}{}_i{}{}^jF_2^{(4)}{}{}_{abcj}+1\leftrightarrow 2\bigg]k^a\zeta^i\nonumber\\
&&-\bigg[32Q_3F_1^{(4)}{}{}^{b}{}{}_i{}{}^{jk}F_2^{(4)}{}{}_{abjk}+32Q_4F_1^{(4)}{}{}_a{}{}^{bjk}F_2^{(4)}{}{}_{bijk}\bigg]k^a\zeta^i\nonumber\\
&&-\bigg[\frac{32}{3}Q_3F_1^{(4)}{}{}_a{}{}^{jkl}F_2^{(4)}{}{}_{ijkl}+\frac{32}{3}Q_4F_1^{(4)}{}{}_i{}{}^{jkl}F_2^{(4)}{}{}_{ajkl}\bigg]k^a\zeta^i\nonumber\\
 \cA_{C^{(1)}C^{(3)}f}&\!\!\!\!\!\sim\!\!\!\!\!&-32i(4Q_1-2Q_3-Q_4)F^{(2)}{}{}^{ci}F^{(4)}{}{}_{abci}f^{ab}+16iQ_3 F^{(2)}{}{}^{ij}F^{(4)}{}{}_{abij}f^{ab}
\eeqa
Here also the first two amplitudes are symmetric under interchanging the particle labels $1,2$, as expected.

\subsection{$p=3$}

Finally, for $p=3$ the amplitudes are
\beqa
\cA_{C^{(0)}C^{(0)}\Phi}&\!\!\!\!\!\sim\!\!\!\!\!&\frac{16}{1!}\bigg[p_1.N.\zeta Q_3+p_2.N.\zeta Q_4\bigg]\bigg(-2F_1^{(1)}{}{}^{\mu}F_2^{(1)}{}{}_{\mu}-2D^\lambda_{}{}_\rho F_1^{(1)}{}{}^{\rho}F_2^{(1)}{}{}_{\lambda}\bigg)\nonumber\\
&&-\bigg[64Q_3F_1^{(1)}{}{}_iF_2^{(1)}{}{}_a+64Q_4F_1^{(1)}{}{}_aF_2^{(1)}{}{}_i\bigg]k^a\zeta^i\nonumber\\
\cA_{C^{(2)}C^{(2)}\Phi}&\!\!\!\!\!\sim\!\!\!\!\!&\frac{256}{3!}\bigg[p_1.N.\zeta (Q_1-Q_3)+p_2.N.\zeta  (Q_2-Q_4)\bigg]F_1^{(3)}{}{}^{abc}F_2^{(3)}{}{}_{abc}\nonumber\\
&&+\frac{16}{3!}\bigg[p_1.N.\zeta Q_3+p_2.N.\zeta Q_4\bigg]\bigg(-2F_1^{(3)}{}{}^{\mu\nu\rho}F_2^{(3)}{}{}_{\mu\nu\rho}-6D^\lambda_{}{}_\rho F_1^{(3)}{}{}^{\rho\mu\nu}F_2^{(3)}{}{}_{\lambda\mu\nu}\bigg)\nonumber\\
&&+\bigg[32(4Q_1-2Q_3-Q_4)F_1^{(3)}{}{}^{ibc}F_2^{(3)}{}{}_{abc}+1\leftrightarrow 2\bigg]k^a\zeta_i\nonumber\\
\cA_{C^{(4)}C^{(4)}\Phi}&\!\!\!\!\!\sim\!\!\!\!\!&\frac{256}{4!}\bigg[p_1.N.\zeta (Q_1-Q_3)+p_2.N.\zeta  (Q_2-Q_4)\bigg]F_1^{(5)}{}{}^{i abcd}F_2^{(5)}{}{}_{i abcd}\nonumber\\
&&-\frac{32}{5!}\bigg[p_1.N.\zeta Q_3+p_2.N.\zeta Q_4\bigg]\bigg( F_1^{(5)}{}{}^{\mu\nu\rho\sigma\delta}F_2^{(5)}{}{}_{\mu\nu\rho\sigma\delta}+5D^\lambda_{}{}_\rho F_1^{(5)}{}{}^{\rho\mu\nu\sigma\delta}F_2^{(5)}{}{}_{\lambda\mu\nu\sigma\delta}\bigg)\nonumber\\
&&+\bigg[\frac{32}{3}(4Q_1-2Q_3-Q_4)F_1^{(5)}{}{}^{bcd}{}{}_i{}{}^jF_2^{(5)}{}{}_{abcdj}+1\leftrightarrow 2  \bigg]k^a\zeta^i\nonumber\\
&&-\bigg[16Q_3F_1^{(5)}{}{}^{bc}{}{}_i{}{}^{jk}F_2^{(5)}{}{}_{abcjk}+16Q_4F_1^{(5)}{}{}_a{}{}^{bcjk}F_2^{(5)}{}{}_{bcijk}\bigg]k^a\zeta^i\nonumber\\
&&-\bigg[\frac{32}{3}Q_3F_1^{(5)}{}{}_a{}{}^{bjkl}F_2^{(5)}{}{}_{bijkl}+\frac{32}{3}Q_4F_1^{(5)}{}{}_{bi}{}{}^{jkl}F_2^{(5)}{}{}_{abjkl}\bigg]k^a\zeta^i\nonumber\\
\cA_{C^{(0)}C^{(2)}f}&\!\!\!\!\!\sim\!\!\!\!\!&-32i(4Q_1-2Q_3-Q_4)F_1^{(1)}{}{}^aF_2^{(3)}{}{}_{abc}f^{bc}+32iQ_3F_1^{(1)}{}{}^iF_2^{(3)}{}{}_{ibc}f^{bc}\\
\cA_{C^{(2)}C^{(4)}f}&\!\!\!\!\!\sim\!\!\!\!\!&-16i(4Q_1-2Q_3-Q_4)F_1^{(3)}{}{}^{cdi}F_2^{(5)}{}{}_{abcdi}f^{ab}+16iQ_3F_1^{(3)}{}{}^{cij}F_2^{(5)}{}{}_{abcij}f^{ab}\nonumber\\
&&+\frac{16i}{3}Q_4F_1^{(3)}{}{}^{ijk}F_2^{(5)}{}{}_{abijk}f^{ab}\nonumber
\eeqa

We have already shown in the previous section that the S-duality Ward identity transforms the amplitude $\cA_{C^{(0)}C^{(2)}f}$ to three other amplitudes, \eg the amplitude of one dilaton, one B-field and one gauge boson, which are consistent with explicit calculations. The amplitude $\cA_{C^{(4)}C^{(4)}\Phi}$ is invariant under the S-duality  Ward identity, the amplitude $\cA_{C^{(0)}C^{(0)}\Phi}$ transforms to the amplitude of two dilatons and one transverse scalar, the amplitude $\cA_{C^{(2)}C^{(2)}\Phi}$transforms to the amplitude of two B-fields and one transverse scalar, and the amplitude $\cA_{C^{(2)}C^{(4)}f}$ transforms to the amplitude of one B-field, one R-R four-form and one gauge bosons. It would be interesting to calculation these amplitudes explicitly and compare them with the S-duality Ward identity predictions.

\section{T-duality Ward identity}

The S-matrix elements in string theory must satisfy the T-duality Ward identity. We now  verify that the amplitudes that we have found in the previous subsections, satisfy the T-duality Ward identity. If one reduces the theory on a circle with coordinate $y$ and if the D$_p$-brane is alone the circle, then after T-duality the brane transforms to the reduction of D$_{p-1}$-brane on the dual circle. The  D$_{p-1}$-brane  is also orthogonal to the dual circle. The $y$ index in D$_p$-brane which is a world-volume index, becomes a transverse index in the T-dual D$_{p-1}$-brane.

For example, consider reduction of the amplitude $\cA_{C^{(1)}C^{(1)}\Phi}$ in \reef{c3c3phi0} when it is orthogonal to the circle, \ie the $y$ index is a transverse coordinate,
\beqa
\cA_{C^{(1)}C^{(1)}\Phi}&\sim&\frac{256}{1!}\bigg[p_1.N.\zeta (Q_1-Q_3)+p_2.N.\zeta  (Q_2-Q_4)\bigg]F_1^{(2)}{}{}^{\ti a}F_2^{(2)}{}{}_{\ti a}\labell{bbb}\\
&&+\frac{16}{2!}\bigg[p_1.N.\zeta Q_3+p_2.N.\zeta Q_4\bigg]\bigg(-16F_1^{(2)}{}{}^{\ti a}F_2^{(2)}{}{}_{\ti a}-4 F_1^{(2)}{}{}^{\ti\tj}F_2^{(2)}{}{}_{\ti\tj}\bigg)\nonumber\\
&&+\bigg[64(4Q_1-2Q_3-Q_4)F_1^{(2)}{}{}^{\ti\tj}F_2^{(2)}{}{}_{a\tj}+64(4Q_2-Q_3-2Q_4)F_1^{(2)}{}{}_{a\tj}F_2^{(2)}{}{}^{\ti\tj}\bigg]k^a \zeta_\ti\nonumber\\
&&+\frac{256}{1!}\bigg[p_1.N.\zeta (Q_1-Q_3)+p_2.N.\zeta (Q_2-Q_4)\bigg]\bigg(F_1^{(2)}{}{}^{ ya}F_2^{(2)}{}{}_{y a }\bigg)\nonumber\\
&&+\frac{16}{2!}\bigg[p_1.N.\zeta Q_3+p_2.N.\zeta Q_4\bigg]\bigg(-16F_1^{(2)}{}{}^{ ya}F_2^{(2)}{}{}_{ ya}-8F_1^{(2)}{}{}^{\ti y}F_2^{(2)}{}{}_{\ti y}\bigg)\nonumber\\
&&+\bigg[64(4Q_1-2Q_3-Q_4)F_1^{(2)}{}{}^{\ti y}F_2^{(2)}{}{}_{ a}{}{}_y+64(4Q_2-Q_3-2Q_4)F_1^{(2)}{}{}_{ a y}F_2^{(2)}{}{}^{\ti}{}{}^y\bigg]k^{ a}\zeta_\ti\nonumber\\
&&+\bigg[64(4Q_1-2Q_3-Q_4)F_1^{(2)}{}{}^{y \tj}F_2^{(2)}{}{}_{ a}{}{}_\tj+64(4Q_2-Q_3-2Q_4)F_1^{(2)}{}{}_{ a \tj}F_2^{(2)}{}{}^{y}{}{}^\tj\bigg]k^{ a}\zeta_y\nonumber
\eeqa
where $\ti,\tj$ are the transverse indices which do not include the $y$-index and $\z_y$ is the polarization of the transverse scalar fields along the $y$-direction. In above, we have used the implicit assumption in the reduction that  fields do not depend on the $y$-direction. The above reduction of D$_0$-brane amplitude should be reproduced by linear T-duality of the reduction of  D$_1$-brane amplitude when it is along the circle. So consider the reduction of the D$_1$-brane amplitudes $\cA_{C^{(0)}C^{(0)}\Phi}$, $\cA_{C^{(2)}C^{(2)}\Phi}$, $\cA_{C^{(0)}C^{(2)}f}$ and $\cA_{C^{(2)}C^{(0)}f}$, \ie  
\beqa
\cA_{C^{(0)}C^{(0)}\Phi}&\!\!\!\!\!\sim\!\!\!\!\!&\frac{256}{1!}\bigg[p_1.N.\zeta (Q_1-Q_3)+p_2.N.\zeta  (Q_2-Q_4)\bigg] F_1^{(1)}{}{}^{\tilde a}F_2^{(1)}{}{}_{\tilde a} \nonumber\\
&&+\frac{16}{1!}\bigg[p_1.N.\zeta Q_3+p_2.N.\zeta Q_4\bigg]\bigg(-8F_1^{(1)}{}{}^{\tilde a}F_2^{(1)}{}{}_{\tilde a} -4F_1^{(1)}{}{}^{i}F_2^{(1)}{}{}_{i}\bigg)\nonumber\\
&&+\bigg[64(4Q_1-2Q_3-Q_4)F_1^{(1)}{}{}^{i}F_2^{(1)}{}{}^{\tilde a}+64(4Q_2-Q_3-2Q_4)F_1^{(1)}{}{}^{\tilde a}F_2^{(1)}{}{}^{i}\bigg]k_{\tilde a}\zeta_i\nonumber\\
\cA_{C^{(2)}C^{(2)}\Phi}&\!\!\!\!\!\sim\!\!\!\!\!&\frac{256}{2!}\bigg[p_1.N.\zeta  (Q_1-Q_3)+p_2.N.\zeta  (Q_2-Q_4)\bigg]\bigg(2F_1^{(3)}{}{}^{iy\tilde a}F_2^{(3)}{}{}_{iy\tilde a}\bigg)\nonumber\\
&&-32\bigg[p_1.N.\zeta Q_3+p_2.N.\zeta Q_4\bigg]\bigg(F_1^{(3)}{}{}^{ijy}F_2^{(3)}{}{}_{ijy} +4F_1^{(3)}{}{}^{iy\tilde a}F_2^{(3)}{}{}_{iy\tilde a} +3F_1^{(3)}{}{}^{y\tilde a\tilde b}F_2^{(3)}{}{}_{y\tilde a\tilde b}\bigg)\nonumber\\
&&+\bigg[64(4Q_1-2Q_3-Q_4)F_1^{(3)}{}{}^{yij}F_2^{(3)}{}{}_{\tilde ayj}+64(4Q_2-Q_3-2Q_4)F_1^{(3)}{}{}_{\tilde ayj}F_2^{(3)}{}{}^{yij}\bigg]k^{\tilde a}\zeta_i\nonumber \\
\cA_{C^{(0)}C^{(2)}f}& \!\!\!\!\!\sim\!\!\!\!\! &64(4Q_1-2Q_3-Q_4)F_1^{(1)}{}{}^iF_2^{(3)}{}{}_{i{\tilde a} y}k^{\tilde a}\z^y\nonumber\\
\cA_{C^{(2)}C^{(0)}f}& \!\!\!\!\!\sim\!\!\!\!\! &64(4Q_2-2Q_4-Q_3)F_2^{(1)}{}{}^iF_1^{(3)}{}{}_{i{\tilde a} y}k^{\tilde a}\z^y\labell{ccc}
\eeqa
where $\tilde a, \tilde b$ are the world-volume indices that do not include the $y$-index and $\z^y$ is polarization of the gauge field along the $y$-direction. In above reduction, we have discarded the $F^{(3)}$-terms that have no $y$-index because they are transformed under T-duality to  $F^{(4)}$-terms that are not included in \reef{bbb}. Using the linear T-duality for the R-R fields, \ie
\beqa
C^{(n)}_{\mu\nu\cdots y}\rightarrow C^{(n-1)}_{\mu\nu\cdots  }&;&C^{(n-1)}_{\mu\nu\cdots}\rightarrow C^{(n)}_{\mu\nu\cdots  y}
\eeqa
and the T-duality for the  gauge field along the circle, \ie
\beqa
A^y\rightarrow \Phi_y
\eeqa
one finds the T-duality of the amplitudes in \reef{ccc} are exactly the amplitude in \reef{bbb}. We have done similar calculations for all other amplitudes and found exact agreement with the T-duality Ward identity.

\section{Soft  scalar theorem}

The S-matrix elements should satisfy the corresponding soft theorems as well. We have proposed in the Introduction section a soft theorem for the scattering amplitude of $n$ closed strings and one soft open string transverse scalar field, \ie equation \reef{A2}. We now check this theorem explicitly for the scattering amplitude of two closed string R-R states and one  scalar field.  

When the open string vertex operator is the transverse scalar, the amplitudes that we have found can be written as 
\beqa
\cA_{2+1}&\sim &\bigg[p_1.N.\zeta (Q_1-Q_3)+p_2.N.\zeta (Q_2-Q_4)\bigg]d_2+\bigg[p_1.N.\zeta Q_3+p_2.N.\zeta Q_4\bigg]d_1\nonumber\\
&&+\hbox{ terms proportional to } k_3\labell{soft1}
\eeqa
where $d_1, d_2$ are
\beqa
d_1&=&\frac{16}{n!} \bigg[\Tr(D)F_1^{(n)}\cdot F_2^{(n)}-2nD^\lambda_{}{}_\rho F_1^{(n)}{}{}^{\rho\mu\cdots\nu}F_2^{(n)}{}{}_{\lambda\mu\cdots\nu}\bigg]\nonumber\\
d_2&=&\frac{256}{n!}\bigg[\delta_{n,p}F_1^{(n)}\cdot V\cdot  F_2^{(n)}+\delta_{n,p+2}F_1^{(n)}{}^{a_0a_1\cdots a_pi}F_2^{(n)}{}_{a_0a_1\cdots a_p i}\bigg]
\eeqa
When the scalar field is soft, \ie  $k_3\rightarrow0$, at the leading order of $k_3$, the terms in the second line of   \reef{soft1} vanish and the integrals in the first line become
\beqa
Q_1-Q_3&=&2\pi B(1+p_1.p_2,p_2.D.p_2)\,{}_{3}F_{2}\bigg[{1+p_1.p_2,\ 0, \  0\atop  1+p_1.p_2+p_2.D.p_2,\ 1}\ ;\ 1
\bigg]\nonumber\\
Q_2-Q_4&=&2\pi B(1+p_1.p_2,p_2.D.p_2+1)\,{}_{3}F_{2}\bigg[{1+p_1.p_2,\ 1, \  1\atop  1+p_1.p_2+p_2.D.p_2+1,\ 1}\ ;\ 1
\bigg]\nonumber\\
Q_3&=&2\pi B(p_1.p_2,p_2.D.p_2+1)\,{}_{3}F_{2}\bigg[{p_1.p_2,\ 0, \  0\atop  p_1.p_2+p_2.D.p_2+1,\ 1}\ ;\ 1
\bigg]\nonumber\\
Q_4&=&2\pi B(p_1.p_2,p_2.D.p_2+2)\,{}_{3}F_{2}\bigg[{p_1.p_2,\ 1, \  1\atop  p_1.p_2+p_2.D.p_2+2,\ 1}\ ;\ 1
\bigg]
\eeqa
Using the following identities:
\beqa
{}_{3}F_{2}\bigg[{a,\ 0, \  0\atop  b,\ c}\ ;\ 1\bigg]&=&1\\
{}_{3}F_{2}\bigg[{a,\ 1, \  1\atop b,\ c}\ ;\ 1\bigg]&=&\frac{\Gamma(c)\Gamma(c-a-b)}{\Gamma(c-a)\Gamma(c-b)}\nonumber
\eeqa
one finds
\beqa
Q_1-Q_3=Q_2-Q_4&= &2\pi B(1+p_1.p_2,p_2.D.p_2)\nonumber\\
Q_3=Q_4&=&2\pi B(p_1.p_2,p_2.D.p_2+1)\labell{soft2}
\eeqa
Replacing them in \reef{soft1}, one finds the amplitude   at the leading order of $k_3$ becomes 
\beqa
\cA_{2+1}&=&\zeta^i(p_1+p_2)_i\cA_2
\eeqa
where
\beqa
\cA_2&\sim&  K(1,2) \frac{\Gamma(p_1.p_2)\Gamma(p_2.D.p_2)}{\Gamma(1+p_1.p_2+p_2.D.p_2)}\labell{AA}
\eeqa
and the kinematic factor $K(1,2)$ is 
\beqa
K(1,2)&=&(p_2.D.p_2d_1+p_1.p_2d_2)
\eeqa
The amplitude \reef{AA} is exactly the disk-level scattering amplitude of two R-R vertex operators \cite{Garousi:2011fc}. So the amplitude that we have found satisfy the soft theorem at the leading order, when the transverse scalar field is soft. It would be interesting to find the subleasing terms in the soft theorem \reef{A2} and check them by explicit calculations.

\section{Soft  photon theorem}

We have proposed in the Introduction section a soft theorem for the scattering amplitude of $n$ closed strings and one soft open string gauge field, \ie equation \reef{A1}. This relation can be used to find the antisymmetric matrix $[\cA_{n}]_{ab}$ that its trace is scattering amplitude of $n$ closed string vertex operators which is zero. Using this theorem, one can find $[\cA_n]_{ab}$ explicitly. 

Using the soft limit of the integrals \reef{soft2}, one finds the matrix $[\cA_2]_{ab}$ has the same structure as the scattering amplitude of two closed string from D-brane, \ie \reef{AA}, with the following kinematic factors:
\beqa
\big[K^{p=1}(C^{(0)},C^{(2)})\big]_{ab}&\!\!\!\!\!\sim\!\!\!\!\!&   p_1\cdot N\cdot p_2F_1^{(1)}{}{}^iF_2^{(3)}{}{}_{iab} \\
\big[K^{p=1}(C^{(2)},C^{(4)})\big]_{ab}&\!\!\!\!\!\sim\!\!\!\!\!&\frac{1}{3!} p_1\cdot V\cdot p_2F_1^{(3)}{}{}_{ijk}F_2^{(5)}{}{}_{ab}{}{}^{ijk}  \nonumber\\
\big[K^{p=2}(C^{(1)},C^{(3)})\big]_{ab}&\!\!\!\!\!\sim\!\!\!\!\!&  p_1\cdot N\cdot p_2F^{(2)}{}{}^{ci}F^{(4)}{}{}_{abci}+\frac{1}{2!} p_1\cdot V\cdot p_2 F^{(2)}{}{}^{ij}F^{(4)}{}{}_{abij} \nonumber\\
\big[K^{p=3}(C^{(0)},C^{(2)})\big]_{ab}&\!\!\!\!\!\sim\!\!\!\!\!&  p_1\cdot N\cdot p_2F_1^{(1)}{}{}^cF_2^{(3)}{}{}_{cab} +   p_1\cdot V\cdot p_2 F_1^{(1)}{}{}^iF_2^{(3)}{}{}_{iab} \nonumber\\
\big[K^{p=3}(C^{(2)},C^{(4)})\big]_{ab}&\!\!\!\!\!\sim\!\!\!\!\!& \frac{1}{2!} p_1\cdot N\cdot p_2F_1^{(3)}{}{}^{cdi}F_2^{(5)}{}{}_{abcdi}+\frac{1}{2!} p_1\cdot V\cdot p_2F_1^{(3)}{}{}^{cij}F_2^{(5)}{}{}_{abcij} \nonumber\\
&& +\frac{1}{3!} p_1\cdot V\cdot p_2F_1^{(3)}{}{}^{ijk}F_2^{(5)}{}{}_{abijk} \nonumber
\eeqa
The trace of the above matrices  are zero. One may use a stringy recursion relation similar to the BCFW field theory recursion relation \cite{Britto:2005fq} to construct the amplitude of four R-R states, \ie $\cA_{C^{(n)}C^{(n)}C^{(n+2)}C^{(n+2)}}$, from  above two-point functions.

{\bf Acknowledgments}:   This work is supported by Ferdowsi University of Mashhad under grant 3/31586(1393/04/30).

 \end{document}